\documentclass{eas-rsk}
\usepackage{graphicx}
\usepackage[comma]{natbib}

\newcommand{\sig}{\lower0.6ex\hbox{$\stackrel{\textstyle >}{\sim}$}\:}
\newcommand{\sil}{\lower0.6ex\hbox{$\stackrel{\textstyle <}{\sim}$}\:}
\newcommand{\sigs}{\lower0.4ex\hbox{$\stackrel{\scriptstyle
      >}{\scriptstyle \sim}$}\,}
\newcommand{\sils}{\lower0.4ex\hbox{$\stackrel{\scriptstyle
      <}{\scriptstyle \sim}$}\,}

\newcommand{\hii}{H$\,${\sc ii}}

\begin{document}

\title{Star Formation in Molecular Clouds} 
\author{Ralf S.\ Klessen}\address{Zentrum f{\"u}r Astronomie der Universit{\"a}t Heidelberg, Institut f{\"u}r Theoretische Astrophysik, Albert-Ueberle-Str. 2, 69120 Heidelberg, Germany}
\begin{abstract}

Stars and star clusters form by gravoturbulent fragmentation of interstellar gas clouds. The supersonic turbulence ubiquitously observed in Galactic molecular gas generates strong density fluctuations with gravity taking over in the densest and most massive regions. Collapse sets in to build up stars. 

Turbulence plays a dual role. On global scales it provides support, while at the same time it can promote local collapse.  Stellar birth is thus intimately linked to the dynamical behavior of parental gas cloud, which governs when and where protostars form, and how they contract and grow in mass via accretion from the surrounding cloud material. The thermodynamic behavior of the star forming gas plays a crucial part in this process and influences the stellar mass function as well as the dynamic properties of the nascent stellar cluster. 

This lecture provides a critical review of our current understanding of stellar birth and compares observational data with competing theoretical models. 

\end{abstract}
\maketitle

\section{Overview}
\label{sec:overview}

When we look at the sky on a clear night, we can note dark patches of obscuration along the band of the Milky Way. These are clouds of dust and gas that block the light from stars further away. Since about one century we know that these clouds are the birthplaces of new stars. With the current set of telescopes and satellites we can observe dark clouds at essentially all wavelengths possible, ranging from $\gamma$-rays to radio frequencies. Especially useful for studying star-forming regions is radio, submillimeter, and far-infrared (IR) emission. We have learned that all star formation occurring in the Milky Way is associated with these dark clouds of molecular hydrogen and dust.
These regions are sufficiently dense and well-shielded against the dissociating effects of interstellar ultraviolet radiation so that hydrogen atoms bind together to form molecules. Molecular hydrogen is a homonuclear molecule, as a consequence its dipole moment vanishes and it radiates extremely weakly. Direct detection of cold interstellar H$_2$ therefore is generally possibly only through ultraviolet (UV) absorption studies. Due to atmospheric opacity these studies can only be done from space, and are limited to pencil-beam measurements of the absorption of light from bright stars or active galactic nuclei. Note that rotational and ro-vibrational emission lines from H$_2$ have also been detected in the infrared, both in the Milky Way and in other galaxies. However this emission comes from gas that has been strongly heated by shocks or radiation, and it traces only a small fraction of the total H$_2$ mass \citep[e.g.][]{vanderwerf00}. Due to these limitations, the most common tool for studying the molecular ISM is radio and sub-millimeter emission either from dust grains or from other molecules that tend to be found in the same locations as H$_2$. The most important tracer molecule is CO, but also HCN, N$_2$H$^+$, NH$_3$ and others are used depending on the density and temperature range of interest. 

Location and mass growth of young stars are  intimately
coupled to the dynamical properties of their parental clouds. Stars form by
gravitational collapse of shock-compressed density fluctuations
generated from the supersonic turbulence ubiquitously observed in
molecular clouds \cite[e.g.][]{maclow04,mckee07}.  
Once a gas clump becomes
gravitationally unstable, it begins to collapse and the central
density increases considerably, giving birth to a protostar. In this
dynamic picture, star formation takes place roughly on a free-fall
timescale, as opposed to the ``standard'' model of the inside-out
collapse of singular isothermal spheres, where core formation is
dominated by the ambipolar diffusion timescale \citep{shu87}. Supersonic turbulence creates a highly  transient and inhomegeneous 
molecular cloud structure which is characterized by large density contrasts. Some the high-density fluctuations 
exceed the critical mass for gravitational
contraction.  The collapse of these  Jeans-unstable cores leads to the formation of individual stars and star clusters.  In this phase, a nascent
protostar grows in mass via accretion from the infalling envelope
until the available gas reservoir is exhausted or stellar feedback
effects become important and remove the parental cocoon --- a new star
is born.  

The structure of these lecture notes is as follows. First, we summarize the global properties of molecular clouds in Section \ref{sec:phenomenology}, then we introduce basic aspects of interstellar turbulence in Section \ref{sec:turbulence}, and turn to the small-scale characteristics of molecular clouds in Section \ref{sec:cores}, where we discuss the properties of the cloud cores that are the direct progenitors of individual stars and binary systems. The statistical properties of stars and star clusters are the focus of Section \ref{sec:stars}, and  we end  with a critical analysis of  different star formation theories in  Section \ref{sec:SF}. By and large, these notes are based on two review articles, \citet{maclow04} and \citet{klessen11}.

\begin{figure}[t]
\begin{center}
\includegraphics[width=0.92\textwidth]{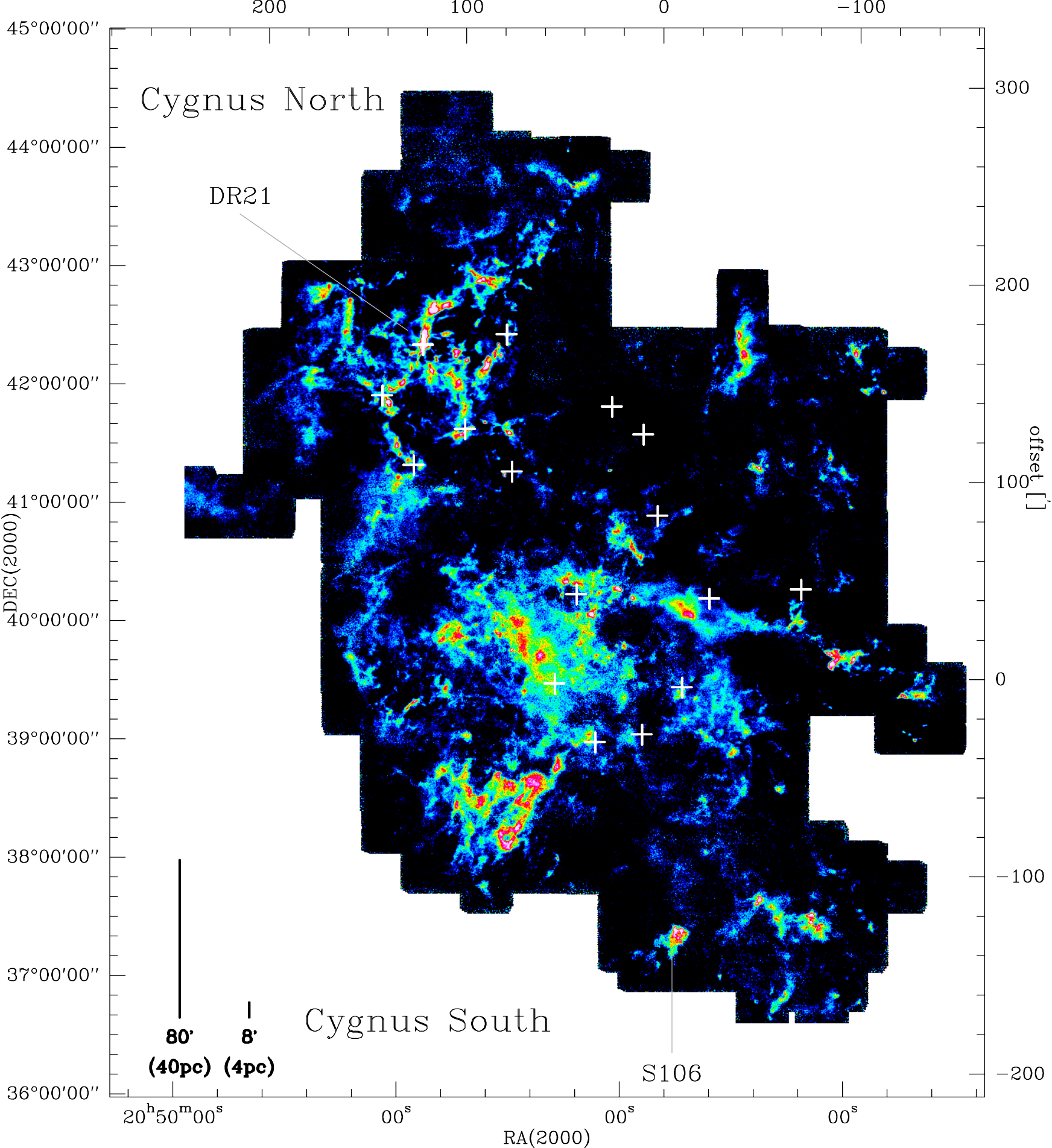}
\caption{The Cygnus X region is the richest and most massive complex of high-mass star formation at a distance lower than $3\,$kpc. \citet{schneider06,schneider07} showed that the Cygnus X region constitutes a large scale network of GMCs at a common distance of $\sim 1.7\,$kpc. Signposts of recent and ongoing (high-mass) star formation (H II regions, outflow activity, IR-sources such as S106 IR, DR21, W75N, and GL2591) are ubiquitous. See  \citet{schneider06} for a review on Cygnus X. The region also contains several OB clusters (Cyg OB1Ð4, 6, and 8) including the richest known OB cluster of the Galaxy, Cyg OB2 \citep{knoedlseder00}. {\em Image from \citet{schneider11}.}}
\label{fig:cygnus}
\end{center}
\end{figure}

\section{Global Properties of Molecular Clouds\label{s:molcloud}}
%
\label{sec:phenomenology}

Molecular clouds in the Milky Way display a number of common properties. First, when studied with high spatial resolution clouds, they exhibit extremely complex and often filamentary structure, with column densities and corresponding 3-D densities that vary by many orders of magnitude. See Figure \ref{fig:cygnus} for an image of Cygnus X, one of the richest and most massive complex of high-mass star formation in the Galaxy. See also Table \ref{tab:MC-prop} for appropriate numbers. We  discuss the small-scale structure and its relation to star formation in Section \ref{sec:cores} below. On large scales, howerver, i.e. when observed at great distance or with low resolution, to within factors of a few all molecular clouds seem to have a similar mean surface density of $\sim 100$ $M_{\odot}$ pc$^{-2}$ corresponding to $0.035$ g cm$^{-2}$ \citep{heyer08, bolatto08a}. The constant surface density of molecular clouds is known as one of 
the Larson \citeyearpar{larson81} relations. We note, however, that there are a number of caveats with these relations and their interpretation \citep{ballesteros02}. Second, the clouds all display linewidths much greater than expected from thermal motion at typical temperatures of $10 - 20$ K. On average, the observed linewidth is related to the size of the cloud by
\begin{equation}
\label{eqn:larson-2}
\sigma_{\rm 1D} = 0.5 \left(\frac{L}{1.0\mbox{ pc}}\right)^{0.5}\mbox{ km s}^{-1},
\end{equation}
where $\sigma_{\rm 1D}$ is the one-dimensional cloud velocity dispersion and $L$ is the cloud size \citep{solomon87, heyer04a, bolatto08a}. This is another one of Larson's relations. The large non-thermal linewidths have been interpreted as indicating the presence of supersonic turbulence, since both the low observed star formation rate (see below) and the absence of inverse P-Cygni line profiles indicates that they are not due to large-scale collapse. If one adopts this interpretation, then from these two observed relations one can directly deduce the third of Larson's relations, which is that giant molecular clouds have virial parameters \citep{bertoldi92}
\begin{equation}
\alpha_{\rm vir} \equiv \frac{5\sigma_{\rm 1D}^2 L}{G M} \approx 1,
\end{equation}
where $M$ is the cloud mass and $G$ is the gravitational constant. This indicates that these clouds are marginally gravitationally bound, but with enough internal turbulence to at least temporarily prevent global collapse.

The presence of supersonic turbulence in approximate virial balance with self-gravity indicates that in molecular clouds the turbulent and gravitational energy densities are of the same order of magnitude, and both generally exceed the thermal energy density by large factors. If molecular clouds form in large-scale convergent flows, as we argue below, then surface terms from ram pressure can also be significant and need to be considered  in the virial equations  \citep{Ballesteros06}. Also magnetic fields are important for the overall energy balance. The gas in molecular clouds is a weakly ionized plasma that is tied to magnetic field lines. Observations using Zeeman splitting \citep{crutcher99, troland08,2010MNRAS.402L..64C} and the Chandrasekhar-Fermi effect \citep{lai01, lai02} indicate that the field strength lies in the range from a few to a few tens of $\mu$G. The exact values vary from region to region, but in general the magnetic energy density appears comparable to the gravitational and turbulent energy densities as expected from energy equipartition arguments. The dynamical importance of the field can be expressed in terms of the magnetic criticality.  If the magnetic field threading a cloud is sufficiently strong, then it cannot undergo gravitational collapse no matter what external pressure is applied to it, as long as it is governed by ideal magnetohydrodynamics (MHD). A cloud in this state is called subcritical. In contrast, a weaker magnetic field can delay collapse, but can never prevent it, and a cloud with such a weak field is called supercritical \citep{1976ApJ...210..326M}. Observations indicate the molecular clouds \citep{padoan99} and cores  \citep{2009ApJ...692..844C,2010MNRAS.402L..64C,2010ApJ...725..466C}, i.e.\ the high-density regions within the clouds that turn into individual stars or small stellar aggregates,  are usually supercritical with values within a factor of a few of the critical one (see also Section \ref{subsec:ind.cores}).

\begin{table}[t]
\begin{center}
{\caption{\label{tab:MC-prop}
 Physical properties of molecular cloud and cores$^a$}}
\begin{tabular}[t]{p{4.5cm}p{1.9cm}p{1.9cm}p{1.9cm}}
\hline
& molecular cloud & cluster-forming clumps & protostellar cores \\
\hline
Size (pc)                            &$2 - 20$   & $0.1-2$    &$\sil 0.1$\\
Mean density (H$_2\, {\rm cm}^{-3}$)   &$10^2-10^3$& $10^3-10^5$&$>10^5$ \\
Mass (M$_{\odot}$)                  &$10^2-10^6$& $10 - 10^3$&$0.1-10$ \\
Temperature (K)                      &$10-30$    & $10-20$    &$7-12$ \\
Line width (km$\,$s$^{-1}$)           &$1 - 10$   & $0.5-3$    &$0.2-0.5$\\
RMS Mach number            &$5 - 50$   & $2 - 15$    &$0 - 2$\\
Column density\\(g cm$^{-2}$) & $0.03$ & $0.03-1.0$ & $0.3-3$ \\
Crossing time (Myr)           & $2 - 10$   & $\sil 1$    &$0.1-0.5$\\
Free-fall time (Myr)           & $0.3 - 3$   & $0.1 - 1$    &$\sil 0.1$\\
Examples                             & Taurus, Ophiuchus & L1641, L1709 & B68, L1544\\
\hline
\end{tabular}
{\footnotesize \vspace{0.2cm} $^a$~Adapted from \citet{Cernicharo1991} and \citet{bergin07}.}
\end{center}
\end{table}

\section{Interstellar Turbulence}
\label{sec:turbulence}
At this point, we need to digress from our discussion of molecular cloud properties and turn our attention to the statistical description of  turbulent flows. In particular, we focus on  the differences between supersonic, compressible (and magnetized) turbulence, which is characteristic of the interstellar medium (ISM), and the more commonly studied incompressible turbulence, which describes terrestrial flows, such as the motion of air in the Earth's atmosphere or the flow of water in rivers and oceans. For the purpose of our discussion, it is sufficient to think of turbulence as the gas flow resulting from random motions at many scales, consistent with the simple scaling relations discussed above. For a more detailed discussion of the complex statistical characteristics of turbulence in general, we refer the reader to the book by \citet{lesieur97}, or for a thorough account of ISM turbulence to the reviews by \citet{elmegreen04} and \citet{Scalo04}.

\subsection{Subsonic, Incompressible Turbulence}
So far, most theoretical studies of turbulence treat incompressible turbulence.
Root-mean-square (rms) velocities are subsonic, and the density remains
almost constant.  Dissipation of energy occurs primarily in the 
smallest vortices, where the dynamical scale $\ell$ is shorter than
the length on which viscosity acts $\eta_{\rm K}$.  \citet{Kolmogorov1941}
described a heuristic theory based on dimensional analysis that
captures the basic behavior of incompressible turbulence surprisingly
well, although subsequent work has refined the details substantially.
He assumed turbulence driven on a large scale $L$, forming eddies at
that scale.  These eddies interact to form slightly smaller eddies,
transferring some of their energy to the smaller scale.  The smaller
eddies in turn form even smaller ones, until energy has cascaded all
the way down to the dissipation scale $\eta_{\rm K}$.  

In order to maintain a steady state, equal amounts of energy must be
transferred from each scale in the cascade to the next, and eventually get 
dissipated, at a rate
\begin{equation}
\dot{E} = \eta v^3/L,
\end{equation}
where $\eta$ is a constant determined empirically. This leads to a
power-law distribution of kinetic energy $E\propto v^2 \propto
k^{-11/3}$, where $k = 2\pi/\ell$ is the wavenumber, and density does
not enter because of the assumption of incompressibility.   The resulting 
differential energy spectrum  is $E(k) dk  \propto k^{-5/3}dk$. Most of the
energy remains near the driving scale, while energy drops off steeply
below $\eta_{\rm K}$.  Because of the apparently local nature of
the cascade in wavenumber space, the viscosity only determines the
behavior of the energy distribution at the bottom of the cascade below
$\eta_{\rm K}$, while the driving only determines the behavior near
the top of the cascade at and above $L$.  The region in between is
known as the inertial range, in which energy gets transported from one scale
to the next without influence from driving or viscosity.  The behavior
of the flow in the inertial range can be studied regardless of the
actual scale at which $L$ and $\eta_{\rm K}$ lie, so long as they
are well separated.  Certain statistical properties of incompressible turbulence, such as 
structure functions $S_p(\vec{r}) = \langle
\{v(\vec{x}) - v(\vec{x}+\vec{r})\}^p \rangle$ for example, have been successfully
modeled by assuming that dissipation occurs in filamentary vortex
tubes \citep{SheLeveque1994}.

Gas flows in the ISM, however, vary from this idealized picture in
three important ways.  First, they are highly compressible, with Mach
numbers ${\cal M}$ ranging from order unity in the warm,
diffuse ISM, up to as high as 50 in cold and dense molecular
clouds.  Second, the equation of state of the gas is very soft due to
radiative cooling, so that pressure $P\propto \rho^{\gamma}$ with the
polytropic index falling in the range $0.4 < \gamma < 1.2$ as a
function of density and temperature \cite[e.g.,][]{scalo98b,ballesterosparedes99,spaans00}. Third, the driving of the turbulence is not uniform, but rather it is inhomogeneous and 
comes from a variety of sources that act on a range of different scales.

\subsection{Supersonic, Compressible Turbulence}
Supersonic flows in highly compressible gas create strong density
perturbations.  Early attempts to understand turbulence in the ISM \cite[e.g.][]{weizsaecker43,Weizsaecker51}
were based on insights drawn from incompressible turbulence.  An attempt to
analytically derive the density spectrum and resulting gravitational
collapse criterion was first made by \citet{1951RSPSA.210...18C,1951RSPSA.210...26C}.  This
work was followed up by several authors, culminating in work by Sasao
(1973) on density fluctuations in self-gravitating media whose
interest has only been appreciated recently.  \citet{larson81} qualitatively applied the basic idea of density
fluctuations driven by supersonic turbulence to the problem of star
formation.
\citet{bonazzola92} used a renormalization group technique to
examine how the slope of the turbulent velocity spectrum could
influence gravitational collapse.
This approach was combined with low-resolution numerical models to
derive an effective adiabatic index for subsonic compressible
turbulence by \citet{panis98}.
Adding to the complexity of the problem, the strong density
inhomogeneities observed in the ISM can be caused not only by
	compressible turbulence, but also by thermal phase transitions \citep{Field69, mckee77, wolfire95}, or gravitational collapse (see, e.g., \citealp{kim01}, for the Galaxy, or \citealp{2010A&A...522A.115S,2010ApJ...721L.134S,2011ApJ...731...62F}, for  applications to high redshift halos).

In supersonic turbulence, shock waves offer additional possibilities
for dissipation.  Shock waves can also transfer energy between widely
separated scales, removing the local nature of the turbulent cascade
typical of incompressible turbulence.  The spectrum may shift only
slightly, however, as the Fourier transform of a step function
representative of a perfect shock wave is $k^{-2}$.  For purely
shock-dominated turbulence, the so called Burgers' turbulence, the 
resulting spectrum is $E(k) dk \propto k^{-2}dk$. Here, certain statistical characteristics can be understood by assuming dissipating occurs in sheetlike shocks \citep{Boldyrev2002, PadoanJimenezNordlundBoldyrev2004,schmidt08}. 
Numerical simulations, 
as well as observations in the interstellar medium, indicate an spectral slope 
that lies somewhere in between Kolmogorov's value of $-5/3$ and the
value of $-2$ for Burgers' turbulence \cite[see, e.g.,][]{federrath09b,federrath10}.

\subsection{Origin of ISM Turbulence}
\label{subsec:origin-turb}
The physical origin of  turbulence in the ISM is not fully understood yet. In particular, the question as to whether it is 
injected from outside or driven by internal sources, i.e.\ whether it is driven on large or small scales,  is still subject to considerable debate. We favor the first assumption as observations indicate that molecular cloud turbulence is always dominated by the largest-scale
modes accessible to the telescope \citep{Ossenkopf02,brunt09}. In addition, the amount of turbulence in molecular clouds with no or extremely low star formation like the Magdalena cloud or the Pipe nebula, is significant and broadly comparable to the level of turbulence observed towards star forming clouds. Both facts seem difficult to reconcile with turbulence being 
driven from internal stellar sources. Instead, we argue that it is the very process of cloud formation that drives its internal 
motions by setting up a turbulent cascade that transports kinetic energy from large to small scales in a universal and 
self-similar fashion.
 
The driving sources for turbulence on the largest scales can be diverse, ranging from the accretion of gas of extragalactic origin \citep{klessen10} to the occurance of convergent flows of atomic gas triggered by spiral density waves \citep{1996A&A...315..265W,1998A&A...330L..21W, 2007ApJ...657..870V, 2009ApJ...707.1023V, 2006ApJ...648.1052H, 2008ApJ...683..786H, 2008A&A...486L..43H, banerjeeetal09}, supernova explosions \citep{maclow04,dib06}, or expanding HII regions \citep{matzner02, krumholz06d, peters08a, gritschneder09a}. Here it is the very process of cloud formation that drives the internal turbulence. Some models have also investigated molecular cloud turbulence that is driven on small scales by internal sources such as stellar winds and outflows \citep{li06b, Nakamura07,wang10} but it is unlikely that these have a significant effect on the largest scales within clouds \citep{maclow04, Banerjee07b, brunt09}.
It is an appealing  hypothesis that molecular clouds form at the stagnation points of large-scale convergent 
flows \citep{BallesterosParedes:1999p13248,BallesterosParedes:2005p13268}. As the density goes 
up the gas can cool efficiently, turn from being mostly atomic to molecular, and shield itself from the external radiation field. 
As long as the convergent flow continues to deliver fresh material the cloud grows in mass and is confined by ram pressure. Recent numerical simulations attempting to form molecular clouds from diffuse gas \citep{2007ApJ...657..870V, 2008ApJ...683..786H, 2008A&A...486L..43H, banerjeeetal09} by considering colliding streams of warm neutral medium, have indeed  shown that this process is sufficient to sustain a substantial degree of  turbulence 
in the forming cloud.

\subsection{Relation to Star Formation}
\label{subsec:relation-to-SF}

\begin{figure}[tbp]
\begin{center}
\includegraphics[width=0.9\columnwidth]{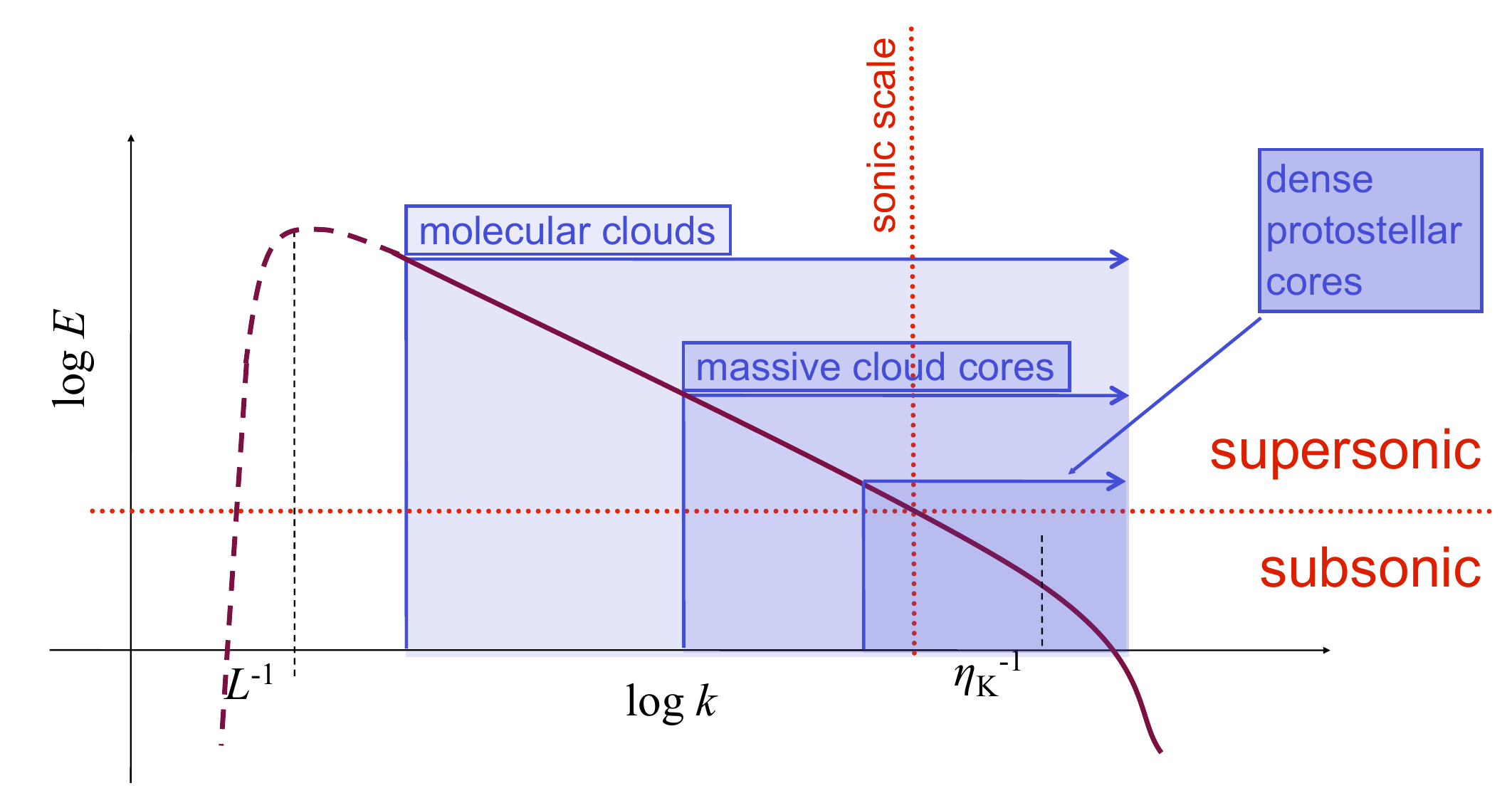}
\caption{Simple cartoon picture of the turbulent energy spectrum, i.e.\ of the kinetic energy carried by modes of different wave numbers $k$, and their relation to different cloud structures (see also Table \ref{tab:MC-prop}). Turbulence is driven on large scales  comparable to the size $L$ of the cloud and is dissipated on very small scales $\eta_{\rm K}$.}
\label{fig:turb-spectrum}
\end{center}
\end{figure}

Altogether, we propose the following picture. On large scales, the turbulence in molecular clouds is highly supersonic. We know that the density contrast for isothermal gas scales with the Mach number $\cal M$ to the second power, that means with ${\cal M} \approx 10$ we expect density contrasts of roughy 100. This is clearly observed in molecular clouds, where the mean density is around $100$ particles per cubic centimeter and where the high-density cores exceed values of $10^4\,$cm$^{-3}$ and more (see Table \ref{tab:MC-prop}). When zooming in on cluster-forming cloud cores (or their not-yet-star-forming counterparts, the so-called infrared dark clouds) one still observes RMS Mach numbers of $\sim 5$, which still leads to localized density fluctuations of a factor of 25 on average. As indicated earlier, some of these fluctuations may exceed the critical mass for gravitational collapse to set in. The presence of turbulence, therefore, makes the cluster-forming core to break apart into smaller units. It fragments to build up a cluster of stars with a wide range of masses rather than forming one single high-mass star. We call this process gravoturbulent fragmentation, because turbulence generates the distribution of clumps initially and then gravity selects a subset of them for star formation. For a more detailed account, see \cite{maclow04}. Finally, when focussing on low-mass cores the velocity field becomes more coherent \cite[e.g.][]{goodman98,bergin07} and the turbulence subsonic. This defines the sonic scale at around $\sim 0.1\,$pc. Such structures are no longer subject to gravoturbulent fragmentation and are the direct progenitors of individual stars or binary systems. We note, however, that gravitational fragmentation may still occur within the protostellar accretion disk that builds up in the center of the core due to angular momentum conservation. This process is thought to produce close binaries \cite[see, e.g.,][]{bodenheimer95,machida08}.  The fact that the observed velocity dispersion approaches the thermal value as one zooms in on smaller and smaller scales, as expressed in Larson's second relation (equation \ref{eqn:larson-2}), is a direct consequence of the negative power-law slope of the turbulent energy spectrum, as illustrated in Figure \ref{fig:turb-spectrum}.

\section{Molecular Cloud Cores}
\label{sec:cores}

In this section focus on the small-scale characteristics of molecular clouds and discuss the properties of the low-mass cores that are the immediate progenitors of individual stars. 

\subsection{Statistical Properties}
\label{sec:statistics}

Emission line observations and dust extinction maps of molecular clouds reveal extremely complex morphological structure with clumps and filaments on all scales accessible by present day telescopes. Typical parameters of different regions in molecular clouds are listed in Table \ref{tab:MC-prop}. The volume filling factor of dense clumps with density contrast $n / \langle n \rangle$, where $n$ is the local density and $\langle n \rangle$ the average density of the cloud, is rather small. It is of order of a few per cent for densities $n >10^5$~cm$^{-3}$ \citep{blitz93,mckee99,williams00,beuther00a}. As discussed above, this is a direct consequence of the fact that ISM turbulence is highly supersonic on large scales.  It is important to note that star formation always occurs in the densest regions within a cloud, so only a small fraction of molecular cloud matter is actually involved in building up stars, while the bulk of the material remains at lower densities. This is the key to explaining the low star formation efficiencies in Galactic molecular clouds.

The mass spectrum of clumps in molecular clouds appears to be well described by a power law,
\begin{equation}
\label{eqn:clump-spectrum}
\frac{dN}{dm} \propto m^{\alpha}\:,
\end{equation}
with the exponent being in the range $-1.3 < \alpha < -1.8$, indicating self-similarity \citep{StutzkiGuesten1990,Williamsetal1994,kramer98a}. There is no natural mass or size scale between the lower and upper limits of the observations. The smallest observed structures are protostellar cores with masses of a few solar masses or less and sizes of $\sil 0.1\,$pc. Given the uncertainties in determining the slope, it appears reasonable to conclude that there is a universal mass spectrum for the clumps within a molecular cloud, and that the distribution is a power law within a mass range of three orders of magnitude, i.e.\ from $1\,$M$_{\odot}$ to about $1000\,$M$_{\odot}$. Hence, it appears plausible that the physical processes that determine the distribution of clump masses are rather similar from cloud to cloud, and that they are closely related to the universal nature of turbulent flows and thermal instability acting on self-gravitating gas. 

Most of the objects that enter in the above morphological analyses are not gravitationally bound \citep{StutzkiGuesten1990,Morata2005,klessen05,Dib2007}. It is interesting to note that the distribution changes as one probes smaller and smaller scales and more and more bound objects. When considering prestellar cores, which are thought to be the direct progenitors of individual stars or small multiple systems, then the mass function is well described by a double power law fit $dN/dm \propto m^{-\alpha}$ following $\alpha = 2.5$ above $\sim0.5\ $M$_{\odot}$ and  $\alpha = 1.5$ below.  The first large study of this kind was published by \citet{motte98}, for a population of submillimetre cores in $\rho$ Oph. Using data obtained with IRAM, they discovered a total of 58 starless clumps, ranging in mass from $0.05\ $M$_{\odot}$ to $\sim 3\ $M$_{\odot}$. Similar results are obtained from the Serpens cloud \citep{testi98}, for  Orion B North  \citep{johnstone01} and Orion B South \citep{Johnstoneetal2006}, or for the Pipe Nebula  \citep{Ladaetal2006}. Currently all observational data \citep{motte98,testi98,johnstone00,johnstone01,Johnstoneetal2006,NutterWardThompson2007,alves07, DiFrancescoetal2007,WardThompsonetal2007,lada08a}  reveal that the mass function of prestellar cores is strikingly similar in shape to the stellar initial mass function, the IMF. To reach complete overlap one is required to introduce a mass scaling or efficiency factor in the range 2 to 10, which differs in different regions. An exciting interpretation of these observations is that we are witnessing the direct formation of the IMF via fragmentation of the parent cloud. However, we note that the observational data also indicate that a considerable fraction of the prestellar cores do not exceed the critical mass for gravitational collapse, much like the clumps on larger scale. The evidence for a one-to-one mapping between prestellar cores and the stellar mass thus is by no means conclusive \cite[see, e.g.,][]{clark07}.

\subsection{Individual Cores}
\label{subsec:ind.cores}

\textit{Density Structure.} 
The density structure of prestellar cores is typically estimated through the analysis of dust emission or absorption using near-IR extinction mapping of background starlight, millimeter/submillimeter dust continuum emission, or dust absorption against the bright mid-IR background emission \citep{bergin07}. A main characteristic of the density profiles derived with the above techniques is that they require a central flattening. The density gradient of a core is flatter than r$^{-1}$ within radii smaller than $2500 - 5000\,$AU, with typical central densities of $10^{5} -10^{6}\,$cm$^{-3}$ \citep{motte98,ward-thompson99}.  A popular approach is to describe these cores as truncated isothermal (Bonnor-Ebert) sphere \citep{ebert55,bonnor56}, that often (but not always) provides a good fit to the data \citep{bacman01,Alves01,kandori05}. These are equilibrium solutions of self-gravitating gas spheres bounded by external pressure. However, such density structure is not unique. Numerical calculations of the dynamical evolution of supersonically turbulent clouds show that the transient cores forming at the stagnation points of convergent flows exhibit similar morphology despite not being in dynamical equilibrium \citep{ballesteros03}.

\begin{figure*}[tbp]
\begin{center}
\includegraphics[height=3.9cm]{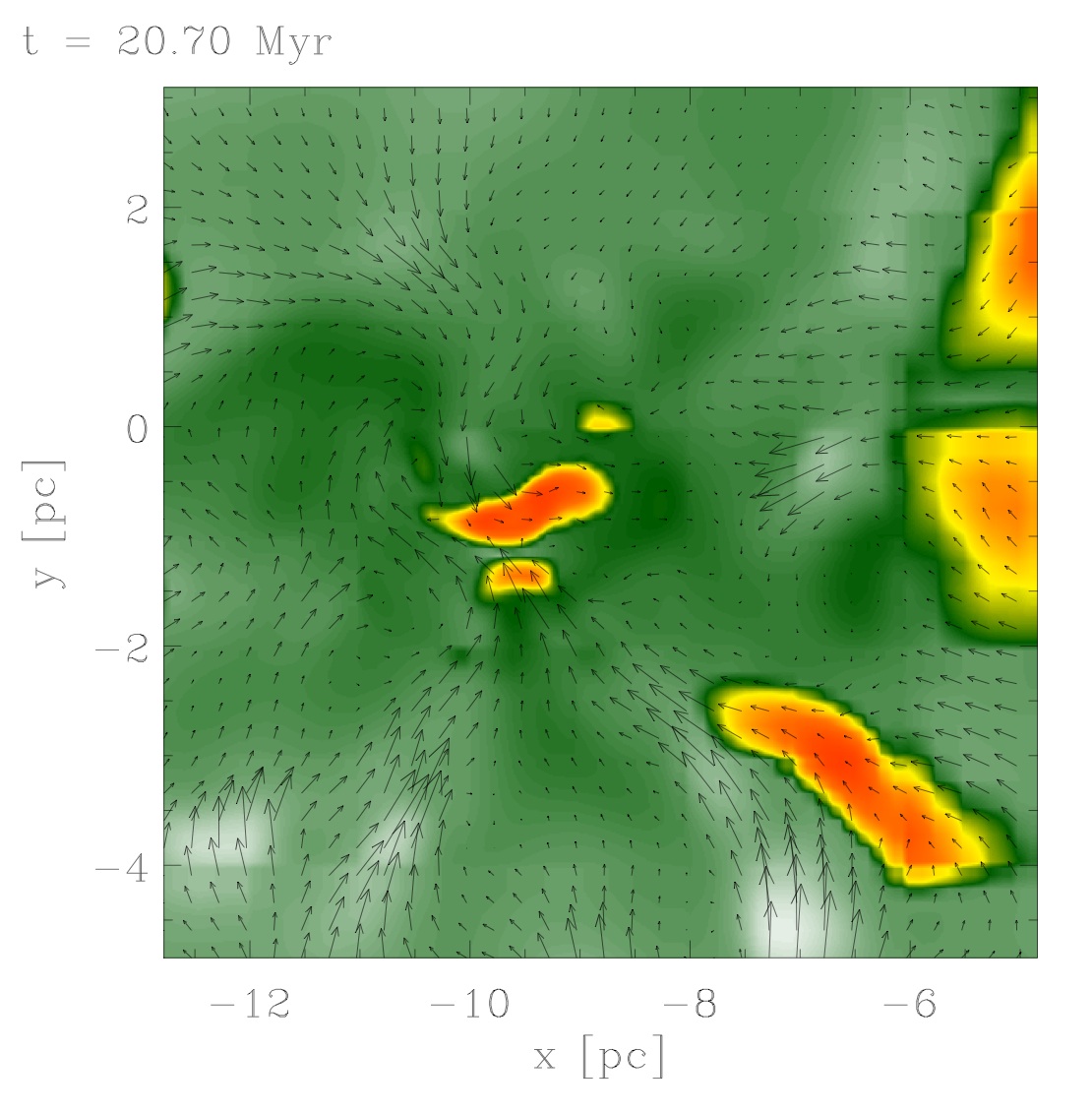}
\includegraphics[height=3.9cm]{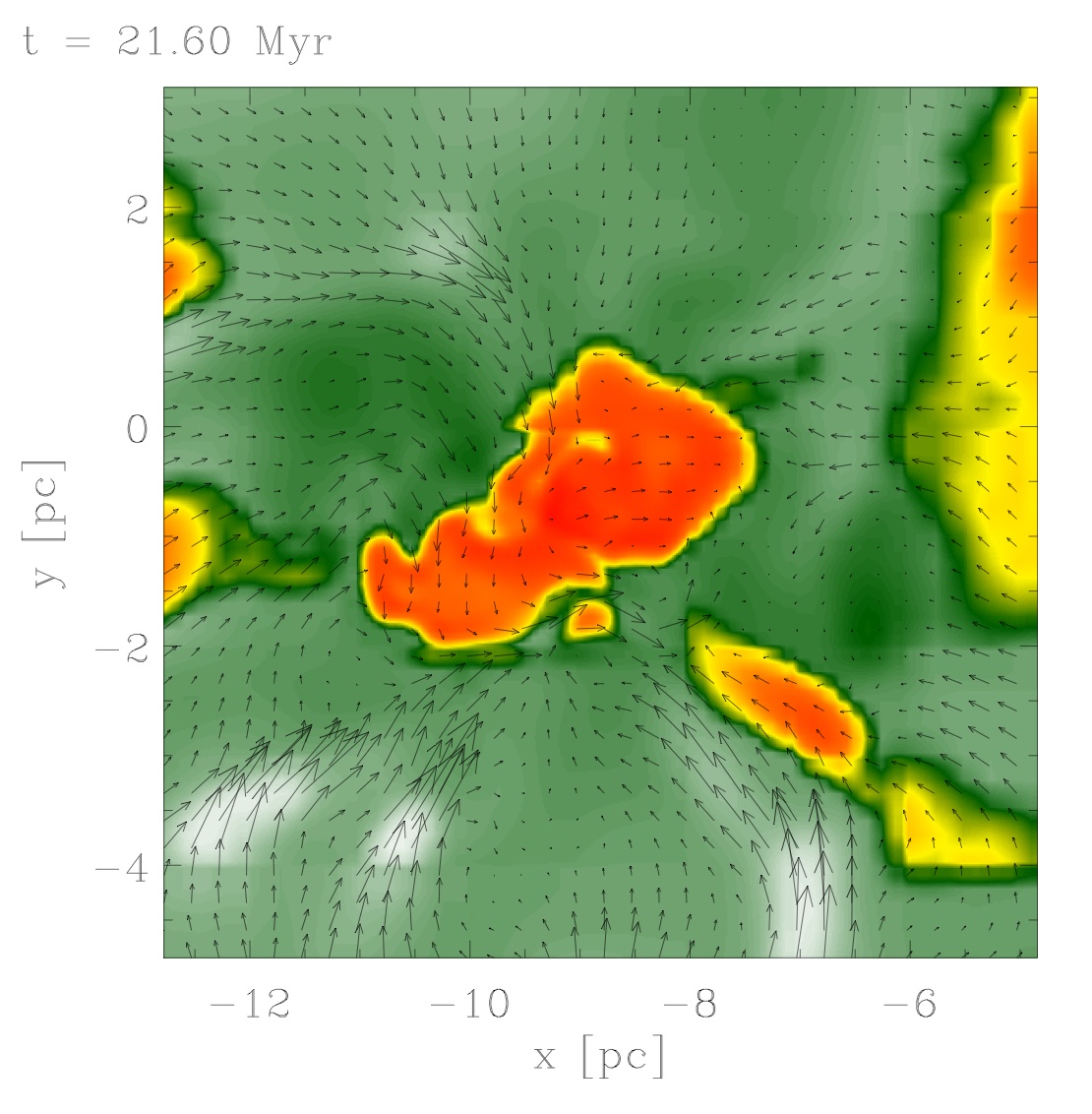}
\includegraphics[height=3.9cm]{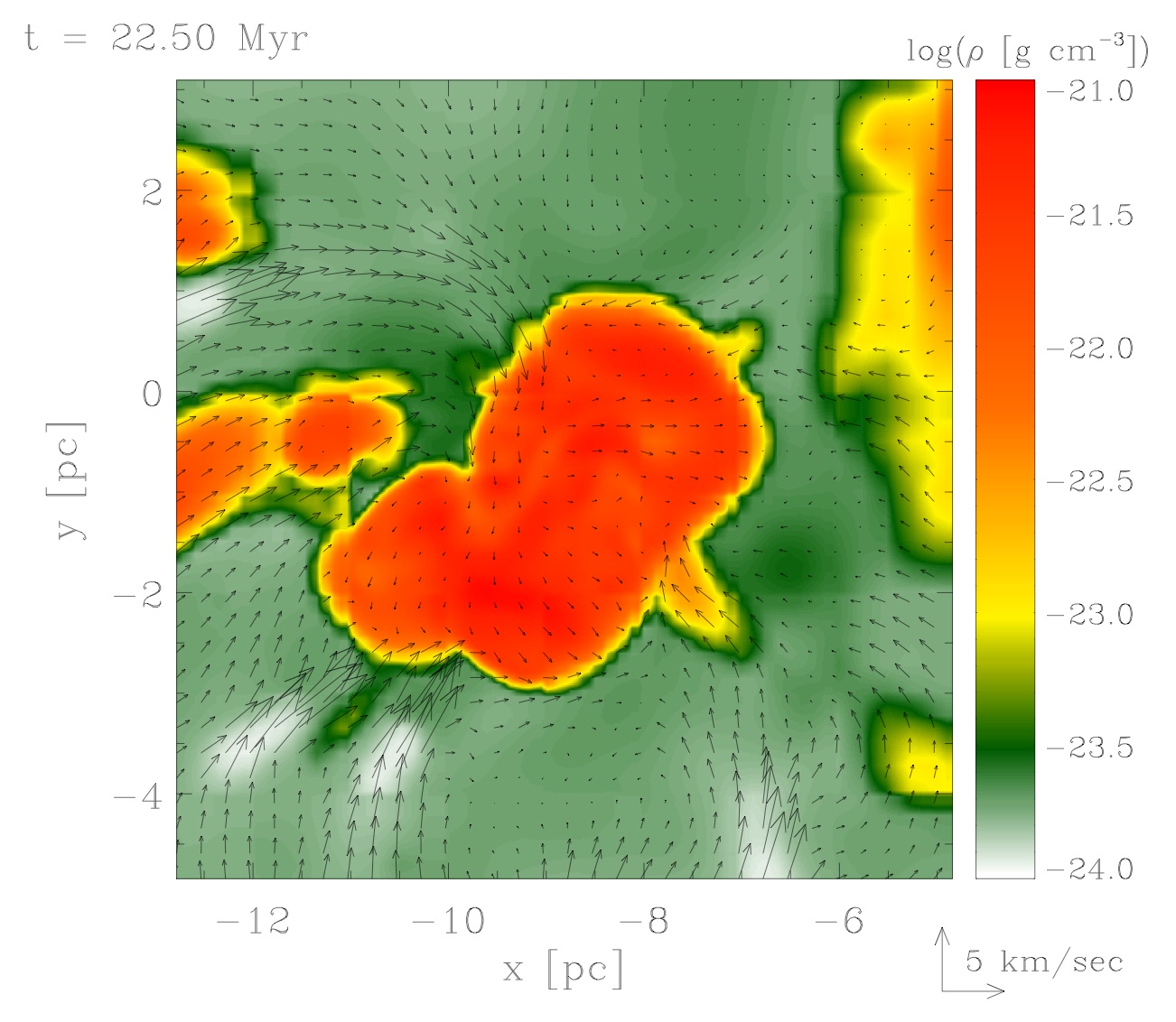}
\caption{Formation and growth of molecular cloud cores by thermal instability triggered by a large-scale convergent flow. A small cold condensate grows from the thermally unstable warm neutral medium by outward propagation of its boundary layer. Coalescence and merging with nearby clumps further increases its mass and size. The global gravitational potential of the proto-cloud enhances the merging probability with time. The images show 2D slices of the density (logarithmic colour scale) and the gas velocity (indicated as arrows) in the plane perpendicular to the large scale flow. {\em From \citet{banerjeeetal09}.}}
\label{fig:clumps}
\end{center}
\end{figure*}

\textit{Thermal Stucture.}
The kinetic temperature of the dust and gas components in a core is  regulated by the interplay between various heating and cooling processes. At high densities ($>10^5$ cm$^{-3}$) in the inner part of the cores, the gas and dust have to be coupled thermally via collisions \citep{goldsmith78,burke83,Goldsmith01}. At lower densities, which correspond to the outer parts of the cores, the two temperatures are not necessary expected to be the same. Thus, the dust and gas temperature distributions need to be inferred from observations independently. Large-scale studies of dust temperature show that the grains in starless cores are colder than in the surrounding lower-density medium. Far-IR observations toward the vicinity of a number of dense cores provide evidence for flat or decreasing temperature gradients with cloud temperatures of $15 -20\,$K and core values of $8 -12\,$K \citep{Ward02,toth04}. These observations are consistent with dust radiative transfer modeling in cores illuminated by interstellar radiation field \citep{langer05,keto05,stamatellos07a}, where the dust temperature is $\sim 7\,$K in the core center and increases up to 16$\,$K in the envelope.  The gas temperature in molecular clouds and cores is commonly infered from the level excitation of simple molecules like CO and NH$_3$ \citep{evans99,walmsley83}. One finds gas temperatures of $10 - 15\,$K, with a possible increase toward the lower density gas near the cloud edges. It is believed that the gas heating in prestellar cores mostly occurs through ionization by cosmic rays, while the cooling is mainly due to line radiation from molecules, especially CO \citep{goldsmith78}. 
Altogether, the fact that prestellar cores are cold and roughly isothermal with at most a modest increase in temperature from the center to the edge is consistent with numerical models of cores forming from thermal instability \citep{heitsch06a,keto08, banerjeeetal09}, see also Figure \ref{fig:clumps}. 

\begin{figure*}[tbp]
\begin{center}
\includegraphics[width=0.95\textwidth]{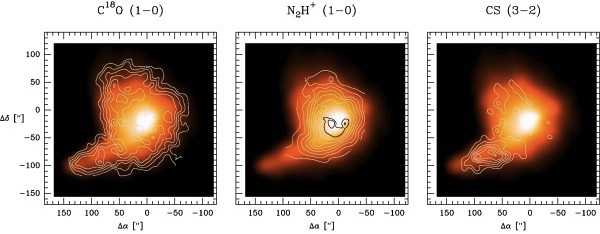}
\caption{Maps of molecular line emission from  C$^{18}$O, N$_2$H$^+$, and CS superimposed on a dust extinction maps of the dark cloud core Barnard 68 \citep{Alves01,bergin02,lada03b}. The three images illustrate the effects of depletion onto grains in the high-density central region of the core. N$_2$H$^+$ is the least and CS the most depleted species. {\em Image courtesy of E.\ A.\ Bergin.}
}
\label{fig:B68}
\end{center}
\end{figure*}

\textit{Chemical Stucture.}
Maps of integral line intensity  can look very different for different molecular tracers. In particular, the N$_2$H$^+$ and NH$_3$ emission more closely follows the dust emission while the C$^{18}$O and CS emission appears as a ``ring-like''\, structure around the dust emission maximum \citep{bergin02,tafalla02,lada03b,maret07a}. For illustration see Figure \ref{fig:B68}. The common theoretical interpretion of these data is that carbon-bearing species, represented by CO and CS freeze-out on the dust grains from the gas while the abundances of nitrogen-hydrogen bearing molecules, N$_2$H$^+$ and NH$_3$, either stay constant or decay more slowly.  At the same time, chemical models of prestellar cores predict that molecules in the core envelope have to be destroyed by interstellar UV field \citep{pavlyuchenkov06,aikawa08}. The chemical stratification significantly complicates the interpretation of molecular line observations and again requires the use of sophisticated chemical models which have to be coupled to the dynamical evolution \cite[see, e.g.,][]{Glover07a,Glover07b,GloverFederrathMacLowKlessen2010}. From observational side, the freeze-out of many molecules makes it difficult to use their emission lines for probing the physical conditions in the inner regions of the cores.  At the same time, the modeling of the chemical evolution can provide us with the important parameters of the cores. For example, the level of CS depletion can be used to constrain the age of the prestellar cores while the deficit of CS in the envelope  can indicate the strength of the external UV field \citep{bergin07}. In any case, any physical interpretation of the molecular lines in prestellar cores has to be based on chemical models and should do justice to the underlying density and velocity pattern of the gas.

\textit{Kinematic Stucture.}
In contrast to the supersonic velocity fields observed in molecular clouds, dense cores have low internal velocities. Starless cores in clouds like Taurus, Perseus, and Ophiuchus systematically present spectra with close-to-thermal linewidths, even when observed at low angular resolution \citep{myers83,jijina99}. This indicates that the gas motions inside the cores are subsonic or at best transsonic, i.e.\ with Mach numbers $\sil 2$  \citep{kirk07, Andreetal2007, rosolowsky08a}. 
In some cores also inward motions have been detected. They are inferred from the observation of optically thick, self-absorbed lines of species like CS, H$_2$CO, or HCO$^+$, in which low-excitation foreground gas absorbs part of the background emission. Typical inflow velocities are of order of $0.05-0.1\,$km/s  and are observed on scales of  $0.05 - 0.15\,$pc, comparable to the observed size of the cores \citep{lee99}. The overall velocity structure of starless cores appears broadly consistent with the structure predicted by models in which protostellar cores form at the stagnation points of convergent flows, but the agreement is not perfect. Simulations of core formation do correctly find that most cores are at most transsonic \citep{klessen05, offner08b}, but the distribution of velocity dispersions has a small tail of highly supersonic cores that is not observed. Clearly more theoretical and numerical work is needed. In particular, the comparison should be based on synthetic line emission maps, which requires to couple a chemical network and radiative transfer to the simulated density profiles as discussed above. In addition, it is also plausible that the discrepancy occurs because the simulations do not include all the necessary physics such as radiative feedback and magnetic fields. Subsonic turbulence contributes less to the energy budget of the cloud than thermal pressure and so cannot provide sufficient support against gravitational collapse \citep{myers83,goodman98,tafalla06}. If cores are longer lasting entities there must be other mechanisms to provide stability. Obvious candidates are magnetic fields \citep{shu87}. However, they are usually not strong enough to provide sufficient support \citep{crutcher99a,crutcher00,bourke01,2009ApJ...692..844C,2010ApJ...725..466C} as discussed below. Most observed cores are thus likely to be evolving transient  objects that never reach any equilibrium state.  

\begin{figure*}[htbp]
\begin{center}
\includegraphics[width=0.90\textwidth]{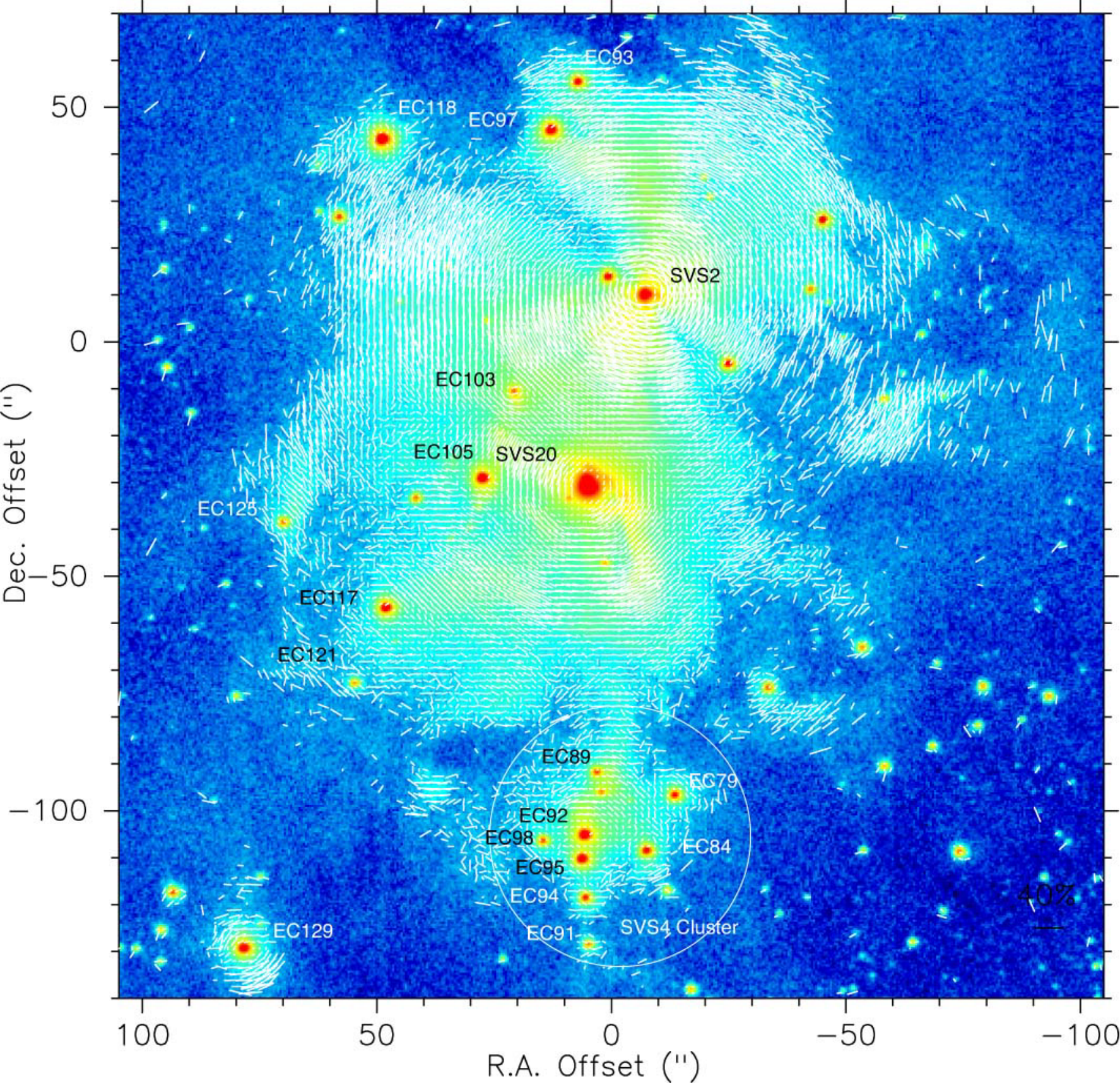}
\caption{Polarization vector map of the central region of the Serpens cloud core, superposed on the total intensity images in logarithmic scaling.  The area of the image is 220'' $\times$ 220''. {\em From \citet{sugitani10}.}
}
\label{fig:mag-serpens}
\end{center}
\end{figure*}

\textit{Magnetic Field Structure.}
Magnetic fields are ubiquitously observed in the interstellar gas on all scales \citep{Crutcher03,heiles05}. However, their importance for star formation and for the morphology and evolution of molecular cloud cores remains controversial. A crucial parameter in this debate is the ratio between core mass and magnetic flux. 
In supercritical cores, this ratio exceeds a critical value and collapse can proceed. In subcritical ones, magnetic fields provide  stability \citep{spitzer78,mouschovias91b,mouschovias91a}. 
Measurements of the Zeeman splitting of molecular lines in nearby cloud cores indicate mass-to-flux ratios that lie above the critical value, in some cases only by a small margin but very often by factors of many  if non-detections are included \citep{crutcher99, bourke01, 2009ApJ...692..844C,2010MNRAS.402L..64C}. The polarization of dust emission offers an alternative pathway to studying the magnetic field structure of molecular cloud cores. MHD simulations of turbulent clouds predict degrees of polarization between 1 and 10\%, regardless of whether turbulent energy dominates over the magnetic energy (i.e.\ the turbulence is super-Alfv\'enic) or not \citep{padoan99,padoan01}. However, converting polarization into magnetic field strength is very difficult \citep{heitsch01b}.  Altogether, the current observational finding imply that magnetic fields must be considered when studying stellar birth, but also that they are not the dominant agent that determines when and where stars form within a cloud. Magnetic fields appear too weak to prevent gravitational collapse to occur.

This conclusion means that in many cases and to reasonable approximation purely hydrodynamic calculations are sufficient for star formation simulations. However, when more precise and quantitative predictions are desired, e.g.\ when attempting to predict star formation timescales or binary properties, it is necessary to perform magnetohydrodynamic (MHD) simulations or even consider non-ideal MHD. The latter means to take ambipolar diffusion (drift between charged and neutral particles) or Ohmic dissipation into account. Recent numerical simulations have shown that even a weak magnetic field can have noticeable dynamical effects. It can alter how cores fragment \citep{price07a, price08a, hennebelle08a, hennebelle08c,hennebelle09,hennebelle11,Peters2011}, change the coupling between stellar feedback processes and their parent clouds \citep{Nakamura07, krumholz07f}, influence the properties of protostellar disks due to magnetic braking \citep{price07b, mellon08b}, or slow down the overall evolution \citep{Heitschetal2001}.

\section{Statistical Properties of Stars and Star Clusters}
\label{sec:stars}

In order to better understand how gas turns into stars, we also need to discuss here some of the key properties of young stellar systems. We restrict ourself to a discussion of the star formation timescale, the spatial distribution of young stars, and the stellar initial mass function (IMF). We note, however, that other statistical characteristics, such as the binary fraction, their relation to the stellar mass, and the orbital parameters of binary stars are equally important for distinguishing between different star formation models.

\subsection{Star Formation Timescales}
The star formation process in molecular clouds appears to be fast \citep{ Hartmann01, Elmegreen07}.
Once the collapse of a cloud region sets in, it rapidly forms an
entire cluster of stars within $10^6$ years or less. This is indicated
by the young stars associated with star forming regions,
typically T~Tauri stars with ages less than $10^6$ years \citep{gomez92,green95, carpenter97}, and by the small age spread in more evolved stellar clusters \citep{Hillenbrand97, palla99}. 
Star clusters in the Milky Way also exhibit an amazing degree of
chemical homogeneity \cite[in the case of the Pleiades, see][]{wilden02}, implying that the gas
out of which these stars formed must have been chemically well-mixed
initially \cite[see also][]{deavillez02, klessen03b}.

\subsection{Spatial Distribution} 
The advent of sensitive infrared detectors in the last decade has made it possible to perform wide-area surveys. These have led us to recognize that most stars form in clusters and aggregates of various size and mass scales, and that isolated or widely distributed star formation is the exception rather than the rule \citep{lada03}. The complex hierarchical structure of molecular clouds (Figure \ref{fig:cygnus}) provides a natural explanation for this finding. 

Star-forming molecular cloud cores vary enormously in size and mass. In small, low-density, clouds stars form with low efficiency, more or less in isolation or scattered around in small groups of up to a few dozen members. Denser and more massive clouds may build up stars in associations and clusters of a few hundred members.  This appears to be the most common mode of star formation in
the solar neighborhood \citep{adams01}. Examples of star formation in small groups and associations are found in the Taurus-Aurigae molecular cloud \citep{hartmann02}. Young stellar groups with a few hundred members form in the Chamaeleon I \citep{persi00} or $\rho$-Ophiuchi \citep{Bontempsetal2001} dark clouds. Each of these clouds is at a distance of about 130 to 160$\,$pc from the Sun.  Like most of the nearby young star forming regions they appear to be associated with a ring-like structure in the Galactic disk called Gould's belt \citep{poeppel97}.\\

\begin{figure}
\center{\includegraphics[width=0.90\textwidth]{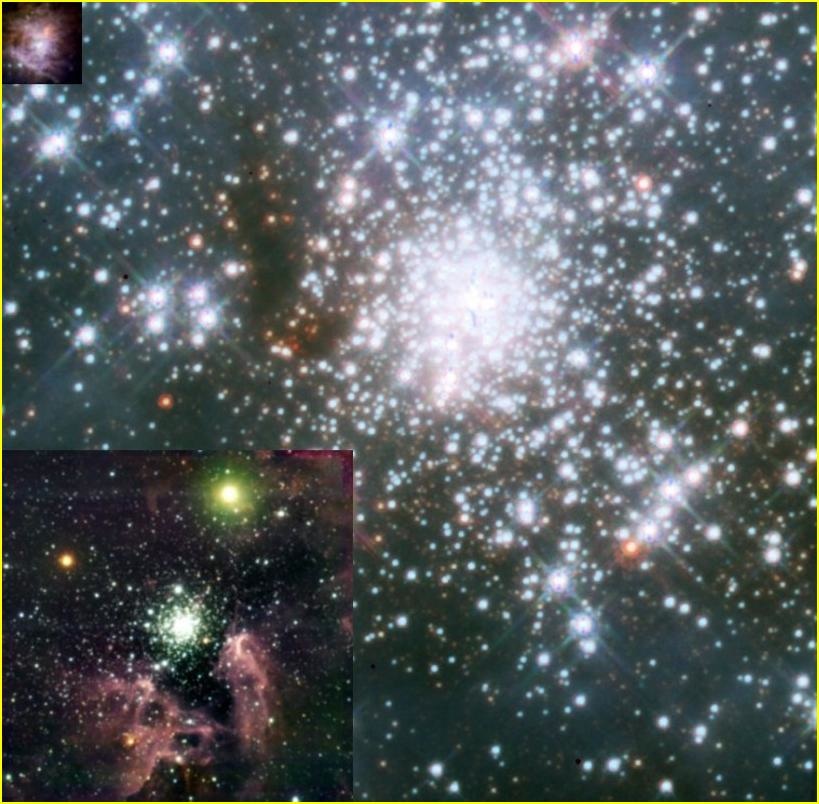}}
\caption{\label{fig:clusters}
Comparison of clusters of different masses scaled to same relative distance. 
The cluster in the upper left corner is the Orion Nebula Cluster \citep{mccaughrean01} and the one at the lower left is NGC~3603 \citep{brandl99}, both observed with the Very Large Telescope at infrared wavelength. The large cluster in the center is 30 Doradus in the LMC observed with the Hubble Space Telescope (courtesy of M.\ J.\ McCaughrean). The total mass increases roughly by a factor of ten from one cluster to the other. {\em Composite image courtesy of H. Zinnecker.}
}
\end{figure}

The formation of dense rich clusters with thousands of stars is rare. The closest region where this happens  is the Orion Nebula Cluster in L1641 \citep{Hillenbrand97,hillenbrand98}, which lies at a distance of $410\,$pc \citep{sandstrom07, menten07, hirota07, caballero08}.  A rich cluster somewhat further away is associated with the Monoceros R2 cloud \citep{carpenter97} at a distance of $\sim 830\,$pc.  The cluster NGC~3603 is roughly ten times more massive than the Orion Nebula Cluster.  It lies in the Carina region, at about $7\,$kpc distance. It contains about a dozen O stars, and is the nearest object analogous to a starburst knot \citep{brandl99,moffat02}. To find star-forming regions building up hundreds of O stars one has to look towards giant extragalactic H\textsc{ii} regions, the nearest of which is 30 Doradus in the Large Magellanic Cloud, a satellite galaxy of our Milky Way at a distance at 55$\,$kpc. The giant star forming region 30 Doradus is thought to contain up to a hundred thousand young stars, including more than 400 O stars \citep{hunter95,walborn97,townsley06}. This sequence as depicted in Figure \ref{fig:clusters} demonstrates that the star formation process spans many orders of magnitude in scale, ranging from isolated single stars to massive young clusters with several $10^4$ stars. This enormous variety of star forming regions appears to be controlled by the competition between self-gravity and opposing agents such as the turbulence in the parental gas clouds, its gas pressure and magnetic field content.

\subsection{The Stellar Initial Mass Function}
\label{subsec:IMF}

Mass is the most important parameter determining the evolution of
individual stars. Massive stars with high pressures at their centers
have strong nuclear fusion, making them short-lived but very
luminous, while low-mass stars are long-lived but extremely faint.
For example, a star with $5\,$M$_{\odot}$ only lives for $2.5\times
10^7\,$yr, while a star with $0.2\,$M$_{\odot}$ survives for
$1.2\times 10^{13}\,$yr, orders of magnitude longer than the current
age of the universe. For comparison the Sun with an age of $4.5\times
10^9\,$yr has reached approximately half of its life span. The
relationship between mass and luminosity is quite steep with roughly $L\propto M^{3.2}$ \citep{kippenhahn94}. During its short life a
$5\,$M$_{\odot}$ star will shine with a luminosity of $1.5\times
10^4\,$L$_{\odot}$, while the luminosity of an $0.2\,$M$_{\odot}$
star is only $\sim10^{-3}\,$L$_{\odot}$. For reference, the luminosity
of the Sun is $1\,$L$_{\odot} = 3.85\times 10^{33}$erg$\,$s$^{-1}$.

\begin{figure}
\center{\includegraphics[width=0.50\textwidth]{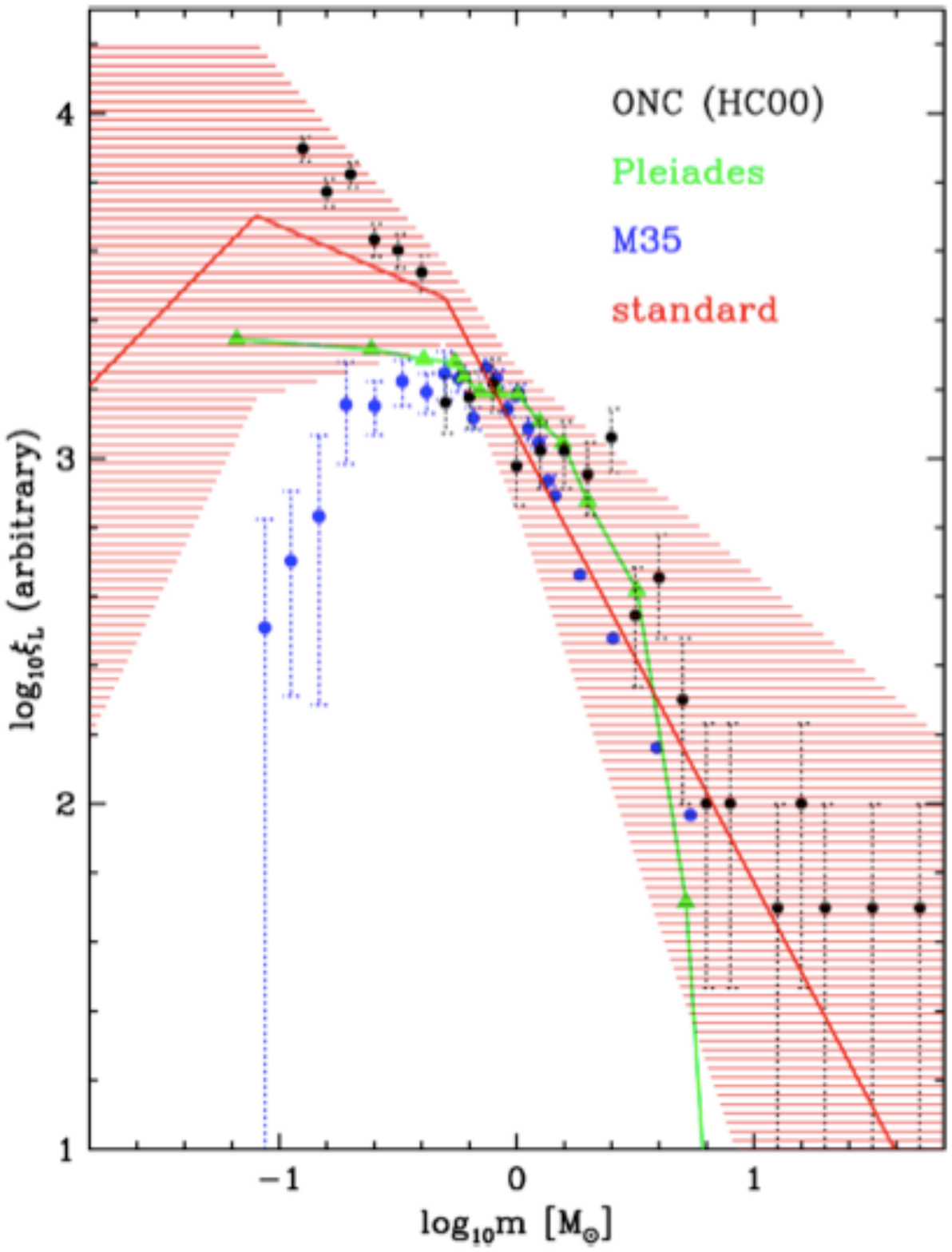}}
\parbox[b]{0.36\textwidth}{\vspace{2.0cm}
\caption{\label{fig:IMF}
Stellar mass spectrum in different nearby clusters (black symbols: Orion Nebula Cluster, green: Pleiades, blue: M35) and its description by a three-component power law (red lines with overall uncertainties indicated by the hatched region). {\em From \citet{kroupa02}.}\vspace*{1.2cm}}} 
\end{figure}

Explaining the distribution of stellar masses at birth, the so-called initial mass function (IMF), is a key prerequisite to any theory of star formation. 
The IMF has three properties that
appear to be relatively robust in diverse environments, see Figure \ref{fig:IMF}. These are the power
law behavior $dN/dm \propto m^{\alpha}$ with slope $\alpha \approx 2.3$ for masses $m$  above about  $1\,$M$_\odot$, originally measured by \citet{Salpeter1955}, the lower mass limit for the power law and the
broad plateau below it before the brown dwarf regime \citep{miller79, scalo86}, and the maximum mass
of stars at around $100\,$M$_\odot$ \citep{weidner04,weidner06,oey05}.
Comprehensive reviews of cluster and field IMFs may be found in \citet{scalo86}, \citet{kroupa02},  \citet{chabrier03}, and \citet{2010ARA&A..48..339B}

At the extreme ends of the stellar mass spectrum, however, our knowledge of the IMF is  limited. Because massive stars are very rare and short lived, only very few are sufficiently near to study them in detail and with very high spatial resolution, for example to determine  multiplicity \citep{zinnecker07}. Low-mass stars and brown dwarfs, on the other hand, are faint, so they too are difficult to study in detail \citep{burrows01}. Such studies, however, are in great demand, because secondary indicators such as the fraction of binaries and higher-order multiples as function of mass, or the distribution disks around very young stars and possible signatures of accretion during their formation are probably better suited to distinguish between different star-formation models than just looking at the IMF. 
In contrast to the observational agreement on the IMF, at least above the substellar regime, there is still considerable disagreement on the theoretical side. The origin of the IMF is a major topic of theoretical research and we can only provide some basic arguments here. For more extended discussions, see \citet{maclow04}, \citet{bonnell07a}, \citet{Larson07}, or \citet{mckee07}.

\section{Star Formation}
\label{sec:SF}

We begin this section with a brief discussion about the possible origin of low-mass stars, which was traditionally the focus on theoretical models of stellar birth. We argue that in the past years, we have seen a paradigm shift away from a slow, quasi-static picture of star formation to a faster and more dynamic approach that is based on the interplay between gravity and turbulence \cite[see also][]{maclow04}. This provides a natural framework for a more consistent understanding of star and star cluster formation. Next we give a critical account of the various proposals in the literature to explain the physical origin of the IMF within the context of the current dynamical theory of star formation. Finally, we turn our attention to the high-mass end of the IMF. Massive stars influence the surrounding universe far out of proportion to their numbers through ionizing radiation, supernova explosions, and heavy element production. Their formation requires the collapse of massive interstellar gas clouds with accretion rates exceeding $10^{-4}$~M$_{\odot}$~yr$^{-1}$ \citep{beutheretal02,beltranetal06} to reach their final masses before exhausting their nuclear fuel \citep{ketoetal06}. For these reasons, the formation of massive stars is less well understood than the formation of low mass stars. However, due to recent advances in algorithmic development and computational capabilities, which allow us to perform 3-dimensional, radiation-magnetohydrodynamic calculations of the collapse and subsequent star formation of massive cloud cores \cite[see, e.g.,][]{klessen11},  our understanding of the origin of the extreme ends of the stellar mass distribution has increased enormously. These developments are discussed in the last part of this Section.

\subsection{Dynamic Star Formation Theory}
The past decade has seen a paradigm shift in low-mass star formation theory. The general believe since the 1980's was that cores in low-mass star-forming regions evolve quasi-statically in magnetically subcritical clouds \citep{shu87}. In this picture, gravitational contraction is mediated by ambipolar diffusion \citep{mouschovias76,mouschovias79,mouschovias81,mouschovias91b} causing a redistribution of magnetic flux until the inner regions of the core become supercritical and go into dynamical collapse.  This process was originally thought to be slow, because in highly subcritical clouds the ambipolar diffusion timescale is about 10 times larger than the dynamical one. However for cores close to the critical value, as is suggested by observations, both timescales are comparable. Numerical simulations furthermore indicate that the ambipolar diffusion timescale becomes significantly shorter for turbulent velocities similar to the values observed in nearby star-forming region  \citep{FatuzzoAdams2002,heitsch04,zli04}. The fact that ambipolar diffusion may not be a slow process under realistic cloud conditions, as well as the fact that most cloud cores are magnetically supercritical \citep{crutcher99a,crutcher00,bourke01,2009ApJ...692..844C}  has cast significant doubts on any magnetically-dominated quasi-static models of stellar birth. For a more detailed account on the shortcomings of the quasi-static, magnetically dominated star formation model, see \cite{maclow04}.

For this reason, star-formation research has turned into considering supersonic turbulence as being on of the primary physical agents regulating stellar birth. The presence of turbulence, in particular of supersonic turbulence, has important consequences for molecular cloud evolution. On large scales it can support clouds against contraction, while on small scales it can provoke localized collapse. Turbulence establishes a complex network of interacting shocks, where dense cores form at the stagnation points of convergent flows. The density can be large enough for gravitational collapse to set in. However, the fluctuations in turbulent velocity fields are highly transient.  The random flow that creates local density enhancements can disperse them again.  For local collapse to actually result in the formation of stars, high density fluctuations must collapse on timescales shorter than the typical time interval between two successive shock passages.  Only then are they able to `decouple' from the ambient flow and survive subsequent shock interactions.  The shorter the time between shock passages, the less likely these fluctuations are to survive. Hence, the timescale and efficiency of protostellar core formation depend strongly on the wavelength and strength of the driving source \citep{klessen00b,Heitschetal2001,Vazquez03,maclow04,krumholz05c,ballesteros07b,mckee07}, and accretion histories of individual protostars are strongly time varying \citep{Klessen2001b,SchmejaKlessen2004}.

\begin{figure}
\center{\includegraphics[width=1.0\textwidth]{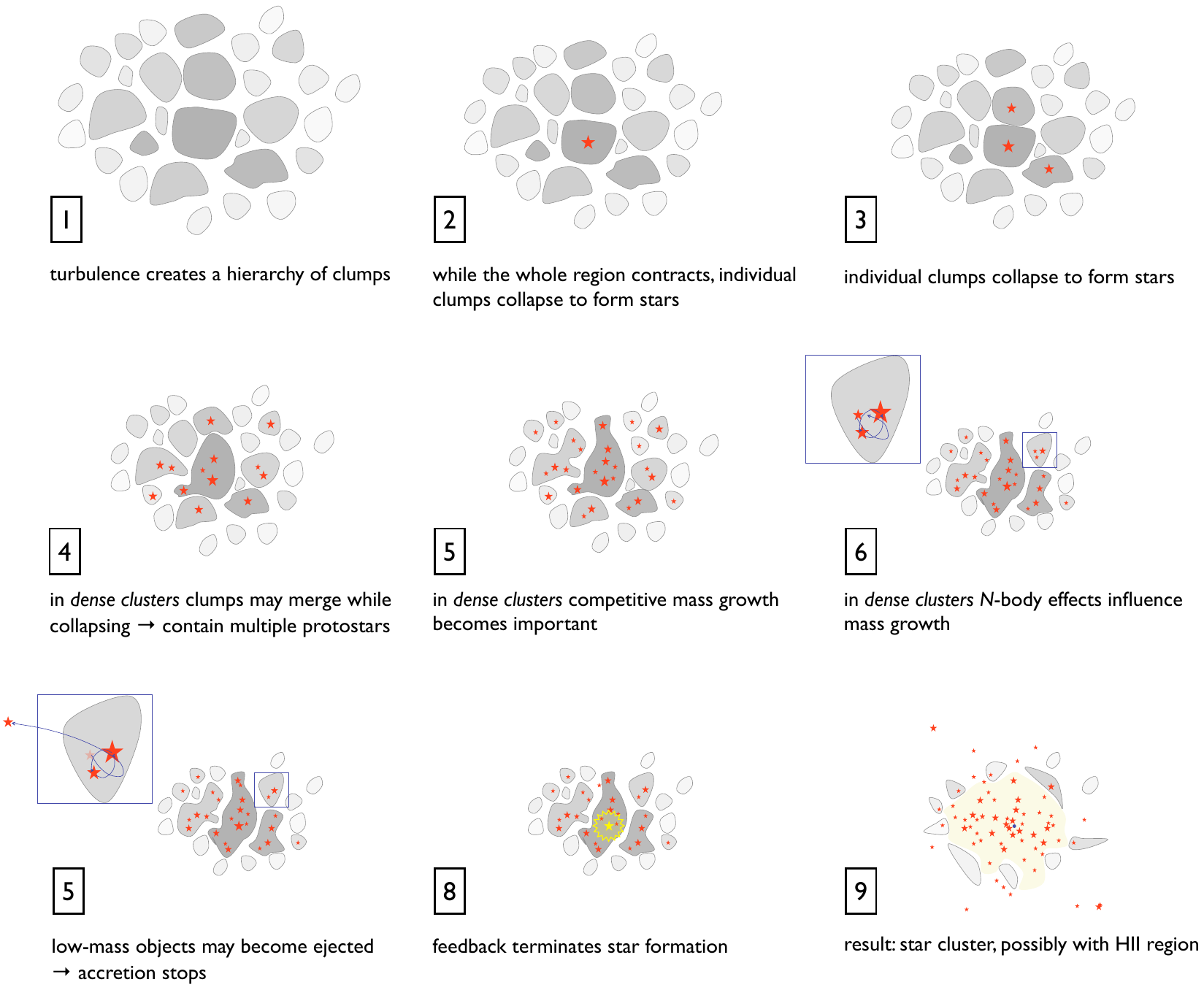}}
\caption{\label{fig:sequence}
Star cluster formation in a turbulent molecular cloud core. }
\end{figure}

Altogether, we propose an evolutionary  sequence as outlined in Figure \ref{fig:sequence}. Following the discussion in Section \ref{subsec:relation-to-SF}, star cluster formation takes place in massive cloud cores of several $10^2$ to $10^3$ solar masses with sizes of a few parsec and a velocity dispersion of $\sim 1\,$km$\,$s$^{-1}$ (see also Table \ref{tab:MC-prop}).  In order to form a bound cluster, the potential energy must dominate the energy budget, meaning that the entire region is contracting. This is a typical feature of the structures formed by large-scale convergent flows (see Section \ref{subsec:origin-turb}). On the scales of a cluster-forming core, the internal turbulent motions are still supersonic with Mach numbers ${\cal M} \sim 5$, and as a consequence there is a high degree of internal substructure with large density contrasts. Some of these density fluctuations are gravitationally unstable in their own right and begin to collapse on timescales much shorter than the global contraction time. Recall that the free-fall time $\tau_{\rm ff}$ scales with the density $\rho$ as $\tau_{\rm ff} \propto \rho^{-1/2}$. 

Typically, the most massive fluctuations have the highest density and form a protostar in their center first. This nascent star can accrete from the immediate environment, but because it is located in a minimum of the cloud core's gravitational potential more gas flows towards it, and it can maintain a high accretion rate for a long time. In contrast, stars that form in lower-mass gas clumps typically can only accrete material from their immediate surrounding and not much beyond that \cite[see, e.g.,][]{klessen00a,klessen01s,klessen01,bonnell04}. Because this preferentially happens in the cluster outskirts, these processes naturally lead to mass segregation as we often observe in young clusters \cite[see, e.g.][ for the Orion Nebula Cluster]{Hillenbrand97,hillenbrand98}. In very dense clusters, there is the possibility that clumps merge while still accreting onto their central protostars. These protostars now compete for further mass growth from a common gas reservoir. This is the essence of the competitive accretion model \citep{Bonnell06}. In its original flavor, it was thought that this leads to the run-away growth of the more massive stars at the expense of the less massive ones \citep{bonnell97,bonnell98}. However, by now we know that the presence of lower-mass companions can significantly reduce the growth rate of massive stars  \citep{Peters2010c}.  This is because neighboring stars will swallow material that would otherwise end up being accreted by the massive stars in the center. We call this process fragmentation-induced starvation and discuss it in more detail in Section \ref{subsec:massive-stars} below. In reality both effects are present and influence the resulting stellar mass spectrum. 

Once a star has reached a mass of $\sim 10\,$M$_{\odot}$ it begins to ionize its environment. It carves out a bubble of hot and tenuous gas, which eventually will expand and enclose the entire stellar cluster. At this point no new stars can form and stellar birth has come to an end. We can observe the young cluster at infrared or even optical wavelength, as illustrated in Figure \ref{fig:clusters}.

\subsection{Models of the Origin of the IMF}
\label{subsec:IMF-models}
There are three dominant schools of thought regarding the origin on the IMF each focusing on different aspects of gravitational collapse in the turbulent ISM. We point out, that the boundaries between these pictures are not clearly defined and that numerous hybrid models have been proposed.

{\textit{Core accretion.}} This model takes as its starting point the striking similarity between the shape of the observed core mass distribution and the IMF. This model assumes a one-to-one relation between the two distributions, such that the observed cores cores are the direct progenitors of individual stars or binary systems. The factor of $\sim\,$3 decrease in mass between cores and stars is thought to be the result of feedback processes, mostly protostellar outflows, that eject a fixed fraction of the mass in a core rather than letting it accrete onto the star \citep{matzner00}. This model reduces the problem of the origin of the IMF to understanding the mass spectrum of bound cores. Arguments to explain the core mass distribution generally rely on the statistical properties of turbulence \citep{Klessen2001, Padoan02, Padoan07, HennebelleChabrier2008,hennebelle09c, chabrier10a,veltchev11}, which generate structures with a pure powerlaw mass spectrum. The thermal Jeans mass in the cloud then imposes the flattening and turn-down in the observed mass spectrum.

{\textit{Competitive accretion.}} A second line reasoning accounts for the fact that stars almost always form in clusters, where the interaction between protostars, and between a protostellar population and the gas cloud around it may become important \citep{bonnell01a, bonnell01b, bonnell02, bate03, bate05, bonnell06d, bonnell08a}. In this picture the origin of the peak in the IMF is much the same as in the core accretion model, it is set by the Jeans mass in the prestellar gas cloud. However, rather than fragmentation in the gas phase producing a spectrum of core masses, each of which collapses down to a single star or star system, in the competitive accretion model all gas fragments down to roughly the Jeans mass. Prompt fragmentation therefore creates a mass function that lacks the powerlaw tail at high masses that we observe in the stellar mass function. This part of the distribution forms via a second phase in which Jeans mass-protostars compete for gas in the center of a dense cluster. The cluster potential channels mass toward the center, so stars that remain in the center grow to large masses, while those that are ejected from the cluster center by $N$-body interactions remain low mass \citep{klessen00a,bonnell04}. In this model, the apparent similarity between the core and stellar mass functions is an illusion, because the observed cores do not correspond to gravitationally bound structures that will collapse to stars \citep{clark06, smith08a}.

{\textit{Importance of the thermodynamic behavior of the gas.}} One potential drawback to both the core accretion and competitive accretion models is that they rely on the Jeans mass to determine the peak of the IMF, but leave unanswered the question of how to compute it. This question is subtle because molecular clouds are nearly isothermal, but they contain a very wide range of densities, and it is unclear which density should be used. A promising idea to resolve this question, which is the basis for a third model of the IMF, focuses on the thermodynamic properties of the gas.  The amount of fragmentation occurring during gravitational collapse depends on the compressibility of the gas \citep{li03}.  For polytropic indices $\gamma < 1$, turbulent compressions cause large density enhancements in which the   Jeans mass falls substantially, allowing many fragments to collapse. Only a few massive fragments get compressed strongly enough to collapse in less compressible gas though. In real molecular gas, the   compressibility varies as the opacity and radiative heating  increase. \citet{Larson05} noted that the thermal coupling of the gas  to the dust at densities  $n_{\rm crit} \sim 10^5\,$cm$^{-3}$  leads to a shift from an adiabatic index of $\gamma \sim 0.7$ to 1.1 as the density increases above $n_{crit}$.  The Jeans mass evaluated at the temperature and density where this shift occurs sets a mass scale for the peak of the IMF. The apparent universality of the IMF in the Milky Way and nearby galaxies may be caused by the insensitivity of the dust temperature on the intensity of the interstellar radiation field \citep{elmegreen08}.  Not only does this mechanism set the peak mass, but also appears to produce a power-law distribution of masses at the high-mass end comparable to the observed distribution \citep{Jappsen05}.

{\textit{Caveats.}} Each of these models has potential problems. In the core accretion picture, hydrodynamic simulations seem to indicate that massive cores should fragment into many stars rather than collapsing monolithically \citep{dobbs05, clark06, Bonnell06}. The hydrodynamic simulations often suffer from over-fragmentation because they do not include radiative feedback from embedded massive stars \citep{krumholz06b,krumholz07a,krumholz08a}. Modeling the formation of massive stars consistently is a very active and challenging field of modern astrophysical research and discussed in more detail in Section \ref{subsec:massive-stars} below. 
The suggestion of a one-to-one mapping between the observed clumps and the final IMF is subject to strong debate, too. Many of the prestellar cores discussed in Section \ref{sec:cores} appear to be stable entities \citep{johnstone00,johnstone01,Johnstoneetal2006,lada08a}, and thus are unlikely to be in a state of active star formation. In addition, the simple interpretation that one core forms on average one star, and that all cores contain the same number of thermal Jeans masses, leads to a timescale problem \citep{clark07} that requires differences in the core mass function and the IMF. 
The criticism regarding neglect of radiative feedback effects also applies to the gas thermodynamic idea. The assumed cooling curves typically ignore the influence of protostellar radiation on the temperature of the gas, which simulations show can reduce fragmentation \citep{krumholz07a}. The competitive accretion picture has also been challenged, on the grounds that the kinematic structure observed in star-forming regions is inconsistent with the idea that protostars have time to interact with one another strongly before they completely accrete their parent cores \citep{Andreetal2007}.

\subsection{Massive Star Formation}
\label{subsec:massive-stars}

Because their formation time is short, of order of $10^5\,$yr, and because they build up deeply embedded in
massive cloud cores, very little is known about the initial and environmental conditions of
high-mass stellar birth. In general high-mass star forming regions are characterized by more
extreme physical conditions than their low-mass counterparts, containing cores of size, mass,
and velocity dispersion roughly an order of magnitude larger than those of cores in low-mass
regions \citep[e.g.][]{jijina99,garliz99,kurtzetal00,beutheretal07,motteetal08}. 
Typical sizes of cluster-forming clumps are $\sim 1\,$pc,
they have mean densities of $n \sim 10^5$ cm$^{-3}$, masses of $\sim 10^3\,$M$_\odot$ and
above, and velocity dispersions ranging between $1.5$ and $4 \,$km$\,$s$^{-1}$. Whenever
observed with high resolution, these clumps break up in even denser cores, that are believed
to be the immediate precursors of single or gravitationally bound multiple massive protostars. 

Massive stars usually form as members of multiple stellar systems \citep{hohaschik81,lada06,zinnecker07}
which themselves are parts of larger clusters \citep{ladalada03,dewitetal04,testietal97}. This
fact
   adds
additional challenges
   to
the interpretation of observational data from high-mass
star forming regions as it is difficult to disentangle mutual dynamical interactions from the
influence of individual stars \citep[e.g.][]{gotoetal06,linzetal05}.
Furthermore, high-mass stars
  reach
the main sequence while still accreting. Their Kelvin-Helmholtz
pre-main sequence contraction
   time
is considerably shorter than
   their accretion time.
Once
  a
star has reached a mass of about $10\,$M$_\odot$ its spectrum becomes UV dominated and it begins
to ionize its environment. This means that accretion as well as ionizing and non-ionizing radiation
needs to be considered in concert \citep{keto02b,keto03,keto07,Peters2010a}. It has been realized
decades ago that in simple 1-dimensional collapse models the outward radiation force on the accreting
material should be significantly stronger than the inward pull of
gravity \citep{larsstarr71,kahn74,wolfcas87}
in particular when taking dust opacities into account. Since there are observations of stars up to $\sim 100\,$M$_\odot$ \citep{bonanosetal04,figer05,rauwetal05} a simple spherically symmetric approach to high-mass star formation must fail. 

Consequently, two different models for massive star formation have been proposed. The first one takes
advantage of the fact that high-mass stars always form as members of stellar clusters. If the central
density in the cluster is high enough, there is a
         chance
that low-mass protostars collide and so
successively build up more massive objects \citep{bonbatezin98}. As the radii of protostars
usually are considerably larger than the radii of main sequence stars in  the same mass range this
could be a viable option. However, the stellar densities required to produce massive stars by collisions are
extremely high \citep{baumgardt11} and seem inconsistent with the observed values of Galactic star clusters 
\cite[e.g.][and references therein]{pozwetal10}.

 An alternative approach is to argue that high-mass stars form like low-mass stars by accretion
of ambient gas that goes through a rotationally supported disk caused by angular momentum conservation.
Indeed such disk structures are observed around a number of high-mass protostars
\citep{chini04,chini06,jiangetal08,daviesetal10}. Their presence
breaks any spherical symmetry that might have been present in the initial cloud and thus solves the
opacity problem. Radiation tends to escape along the polar axis, while matter is transported inwards
through parts of the equitorial plane shielded by the disk. Hydrodynamic simulations in two and three
dimensions focusing on the transport of non-ionizing radiation strongly
support this picture \citep{yorke02,krumholzetal09, kuiper10, kuiper11}.
         The same holds when taking the  effects of ionizing radiation into account \citep{Peters2010a, Peters2010b, Peters2010c, Peters2011}. If the disk becomes          gravitationally unstable, material flows along dense, opaque filaments whereas the radiation
escapes through optically thin channels in and above the disk. Even
ionized material can be accreted, if the accretion flow is strong enough. \hii\ regions are gravitationally
trapped at that stage, but soon begin to rapidly fluctuate between trapped and extended states, in
agreement with observations \citep{Peters2010a,galvan-madrid11}. Over time, the same
ultracompact \hii\ region can expand anisotropically, contract again, and take on any of the observed
morphological classes \citep{woodchurch89,kurtzetal94, Peters2010b}. In their
extended phases, expanding \hii\  regions drive bipolar neutral outflows characteristic of high-mass
star formation \citep{Peters2010a}. 

Another key fact that any theory of massive star formation must account for is the apparent presence of an upper mass limit at around $100 - 150\,$M$_\odot$ \citep{massey03}.
It holds for the Galactic field, however, it is also true for young star clusters
that are massive enough so that purely random sampling of the initial mass function (IMF)
\citep{kroupa02,chabrier03} without upper mass limit should have yielded stars above
$150\,$M$_\odot$ (\citealt{weidner04,figer05,oey05,weidetal10}, see however, \citealt{selmel08}).
This immediately raises the question
of what is the physical origin of this apparent mass limit. It has been speculated before that
      radiative
stellar feedback might be responsible for this limit \citep[for a detailed discussion see, e.g.,][]{zinnecker07} or
alternatively that the internal stability limit of stars with non-zero metallicity lies in this
mass regime \citep{appen70a,appen70b,appen87,baraetal01}.
However,
    fragmentation could also
limit protostellar mass growth, as suggested by the numerical simulations of \citet{Peters2010c}. The likelihood of fragmentation to occur and the number of fragments
to form depends sensitively on the physical conditions in the star-forming cloud and its initial and
environmental parameters \citep[see, e.g.,][]{girichidis11a}. Understanding the build-up of massive
stars, therefore, requires detailed knowledge about the physical processes that initiate and regulate the
formation and dynamical evolution of the molecular clouds these stars form in 
\citep{vazsemetal09}.

\citet{Peters2010c,Peters2011} argue that ionizing radiation, just like its non-ionizing, lower-energy counterpart,
         cannot
shut off the accretion flow onto massive stars. Instead it is the dynamical processes
in the gravitationally unstable accretion
         flow
that inevitably
      occurs
during the collapse of
high-mass cloud cores that
         control
the mass growth of individual protostars.  Accretion onto the
central star is shut off by the fragmentation of the disk and the formation of  lower-mass companions
which intercept inward moving material. They call this process fragmentation-induced starvation and
show that it occurs unavoidably in regions of high-mass star formation where the mass flow onto the
disk exceeds the inward transport of matter due to viscosity only and thus renders the disk unstable
to fragmentation. 

As a side note, it is interesting to speculate that  fragmentation-induced starvation is important not only for present-day star formation but also in the primordial universe during the formation of metal-free Population III stars. Consequently, we expect these stars to be in binary or small number multiple systems and to be of lower mass than usually inferred \citep{abeletal02,brommetal09}. Indeed, current numerical simulations provide the first hints that this might be the case \citep{clark11a,greif11b}.

\section*{Acknowledgements}
I am grateful for may stimulating discussions with Fabian Heitsch, Mark Krumholz, and Mordecai-Mark Mac~Low, who are the co-authors of the review articles this lecture is based on. I also thank Robi Banerjee, Paul Clark, Christoph Federrath, Simon Glover, Patrick Hennebelle, Thomas Peters, Dominik Schleicher, Rahul Shetty, and Rowan Smith for  very fruitful collaborations and for the continuous exchange of scientific ideas that helped to shape the view on stellar birth presented here. I also acknowledge funding from the {\em Baden-W\"urttemberg-Stiftung} (grant P-LS-SPII/18) and from the German {\em Bundesministerium f\"ur Bildung und Forschung} via the ASTRONET project STAR FORMAT (grant 05A09VHA). Further support comes from the {\em Deutsche Forschungsgemeinschaft} under grants no.\ KL 1359/5, KL 1358/10,  KL 1358/11, and the {\em Sonderforschungsbereich} SFB 881 The Milky Way System.

{\setlength{\bibsep}{0.01cm}
\bibliographystyle{astron}
\bibliography{aussois-bib}

\begin{thebibliography}{}

\bibitem[\protect\astroncite{Abel et~al.}{2002}]{abeletal02}
Abel, T., Bryan, G.~L., and Norman, M.~L.: 2002,
\newblock {\em {Science}} {\bf 295}, 93

\bibitem[\protect\astroncite{{Adams} and {Myers}}{2001}]{adams01}
{Adams}, F.~C. and {Myers}, P.~C.: 2001,
\newblock {\em \apj} {\bf 553}, 744

\bibitem[\protect\astroncite{{Aikawa} et~al.}{2008}]{aikawa08}
{Aikawa}, Y., {Wakelam}, V., {Garrod}, R.~T., and {Herbst}, E.: 2008,
\newblock {\em \apj} {\bf 674}, 984

\bibitem[\protect\astroncite{{Alves} et~al.}{2007}]{alves07}
{Alves}, J., {Lombardi}, M., and {Lada}, C.~J.: 2007,
\newblock {\em \aap} {\bf 462}, L17

\bibitem[\protect\astroncite{{Alves} et~al.}{2001}]{Alves01}
{Alves}, J.~F., {Lada}, C.~J., and {Lada}, E.~A.: 2001,
\newblock {\em \nat} {\bf 409}, 159

\bibitem[\protect\astroncite{{Andr{\'e}} et~al.}{2007}]{Andreetal2007}
{Andr{\'e}}, P., {Belloche}, A., {Motte}, F., and {Peretto}, N.: 2007,
\newblock {\em \aap} {\bf 472}, 519

\bibitem[\protect\astroncite{Appenzeller}{1970a}]{appen70b}
Appenzeller, I.: 1970a,
\newblock {\em \aap} {\bf 9}, 216

\bibitem[\protect\astroncite{Appenzeller}{1970b}]{appen70a}
Appenzeller, I.: 1970b,
\newblock {\em \aap} {\bf 5}, 355

\bibitem[\protect\astroncite{Appenzeller}{1987}]{appen87}
Appenzeller, I.: 1987,
\newblock in H.~J.~G.~L.~M. Lamers and C.~W.~H. de~Loore (eds.), {\em
  {Instabilities in Luminous Early Type Stars}}, Vol. 136 of {\em {Astrophysics
  and Space Science Library}}, pp 55--67

\bibitem[\protect\astroncite{{Bacmann} et~al.}{2001}]{bacman01}
{Bacmann}, A., {Andr{\'e}}, P., and {Ward-Thompson}, D.: 2001,
\newblock in T. {Montmerle} and P. {Andr{\'e}} (eds.), {\em From Darkness to
  Light: Origin and Evolution of Young Stellar Clusters}, Vol. 243 of {\em
  Astronomical Society of the Pacific Conference Series}, p. 113

\bibitem[\protect\astroncite{{Ballesteros-Paredes}}{2006}]{Ballesteros06}
{Ballesteros-Paredes}, J.: 2006,
\newblock {\em \mnras} {\bf 372}, 443

\bibitem[\protect\astroncite{Ballesteros-Paredes
  et~al.}{1999}]{BallesterosParedes:1999p13248}
Ballesteros-Paredes, J., Hartmann, L., and V{\'a}zquez-Semadeni, E.: 1999,
\newblock {\em ApJ} {\bf 527}, 285

\bibitem[\protect\astroncite{{Ballesteros-Paredes}
  et~al.}{2007}]{ballesteros07b}
{Ballesteros-Paredes}, J., {Klessen}, R.~S., {Mac Low}, M.-M., and
  {Vazquez-Semadeni}, E.: 2007,
\newblock in B. {Reipurth}, D. {Jewitt}, and K. {Keil} (eds.), {\em Protostars
  and Planets V}, pp 63--80

\bibitem[\protect\astroncite{{Ballesteros-Paredes}
  et~al.}{2003}]{ballesteros03}
{Ballesteros-Paredes}, J., {Klessen}, R.~S., and {V{\'a}zquez-Semadeni}, E.:
  2003,
\newblock {\em \apj} {\bf 592}, 188

\bibitem[\protect\astroncite{{Ballesteros-Paredes} and {Mac
  Low}}{2002}]{ballesteros02}
{Ballesteros-Paredes}, J. and {Mac Low}, M.-M.: 2002,
\newblock {\em \apj} {\bf 570}, 734

\bibitem[\protect\astroncite{Ballesteros-Paredes
  et~al.}{2005}]{BallesterosParedes:2005p13268}
Ballesteros-Paredes, J., V{\'a}zquez-Semadeni, E., and Kim, J.: 2005,
\newblock {\em Protostars and Planets V} p. 8630

\bibitem[\protect\astroncite{{Ballesteros-Paredes}
  et~al.}{1999}]{ballesterosparedes99}
{Ballesteros-Paredes}, J., {V{\'a}zquez-Semadeni}, E., and {Scalo}, J.: 1999,
\newblock {\em \apj} {\bf 515}, 286

\bibitem[\protect\astroncite{{Banerjee} et~al.}{2007}]{Banerjee07b}
{Banerjee}, R., {Klessen}, R.~S., and {Fendt}, C.: 2007,
\newblock {\em \apj} {\bf 668}, 1028

\bibitem[\protect\astroncite{Banerjee et~al.}{2009}]{banerjeeetal09}
Banerjee, R., V{\'a}zquez-Semadeni, E., Hennebelle, P., and Klessen, R.~S.:
  2009,
\newblock {\em \mnras} {\bf 398}, 1082

\bibitem[\protect\astroncite{Baraffe et~al.}{2001}]{baraetal01}
Baraffe, I., Heger, A., and Woosley, S.~E.: 2001,
\newblock {\em \apj} {\bf 550}, 890

\bibitem[\protect\astroncite{{Bastian} et~al.}{2010}]{2010ARA&A..48..339B}
{Bastian}, N., {Covey}, K.~R., and {Meyer}, M.~R.: 2010,
\newblock {\em \araa} {\bf 48}, 339

\bibitem[\protect\astroncite{{Bate} and {Bonnell}}{2005}]{bate05}
{Bate}, M.~R. and {Bonnell}, I.~A.: 2005,
\newblock {\em \mnras} {\bf 356}, 1201

\bibitem[\protect\astroncite{{Bate} et~al.}{2003}]{bate03}
{Bate}, M.~R., {Bonnell}, I.~A., and {Bromm}, V.: 2003,
\newblock {\em \mnras} {\bf 339}, 577

\bibitem[\protect\astroncite{{Baumgardt} and {Klessen}}{2011}]{baumgardt11}
{Baumgardt}, H. and {Klessen}, R.~S.: 2011,
\newblock {\em \mnras} pp 321--+

\bibitem[\protect\astroncite{Beltr{\'a}n et~al.}{2006}]{beltranetal06}
Beltr{\'a}n, M.~T., Cesaroni, R., Codella, C., Testi, L., Furuya, R.~S., and
  Olmi, L.: 2006,
\newblock {\em \nat} {\bf 443}, 427

\bibitem[\protect\astroncite{{Bergin} et~al.}{2002}]{bergin02}
{Bergin}, E.~A., {Alves}, J., {Huard}, T., and {Lada}, C.~J.: 2002,
\newblock {\em \apjl} {\bf 570}, L101

\bibitem[\protect\astroncite{{Bergin} and {Tafalla}}{2007}]{bergin07}
{Bergin}, E.~A. and {Tafalla}, M.: 2007,
\newblock {\em \araa} {\bf 45}, 339

\bibitem[\protect\astroncite{{Bertoldi} and {McKee}}{1992}]{bertoldi92}
{Bertoldi}, F. and {McKee}, C.~F.: 1992,
\newblock {\em \apj} {\bf 395}, 140

\bibitem[\protect\astroncite{Beuther et~al.}{2007}]{beutheretal07}
Beuther, H., Churchwell, E.~B., McKee, C.~F., and Tan, J.~C.: 2007,
\newblock in B. Reipurth, D. Jewitt, and K. Keil (eds.), {\em {Protostars and
  Planets V}}, pp 165--180, {The University of Arizona Press}

\bibitem[\protect\astroncite{{Beuther} et~al.}{2000}]{beuther00a}
{Beuther}, H., {Kramer}, C., {Deiss}, B., and {Stutzki}, J.: 2000,
\newblock {\em \aap} {\bf 362}, 1109

\bibitem[\protect\astroncite{Beuther et~al.}{2002}]{beutheretal02}
Beuther, H., Schilke, P., Sridharan, T.~K., Menten, K.~M., Walmsley, C.~M., and
  Wyrowski, F.: 2002,
\newblock {\em \aap} {\bf 383}, 892

\bibitem[\protect\astroncite{{Blitz}}{1993}]{blitz93}
{Blitz}, L.: 1993,
\newblock in {\em Protostars and Planets III}, pp 125--161

\bibitem[\protect\astroncite{{Bodenheimer}}{1995}]{bodenheimer95}
{Bodenheimer}, P.: 1995,
\newblock {\em \araa} {\bf 33}, 199

\bibitem[\protect\astroncite{{Bolatto} et~al.}{2008}]{bolatto08a}
{Bolatto}, A.~D., {Leroy}, A.~K., {Rosolowsky}, E., {Walter}, F., and {Blitz},
  L.: 2008,
\newblock {\em \apj} {\bf 868}, 948

\bibitem[\protect\astroncite{{Boldyrev}}{2002}]{Boldyrev2002}
{Boldyrev}, S.: 2002,
\newblock {\em \apj} {\bf 569}, 841

\bibitem[\protect\astroncite{Bonanos et~al.}{2004}]{bonanosetal04}
Bonanos, A.~Z., Stanek, K.~Z., Udalski, A., Wyrzykowski, L., {\.Z}ebru{\'n},
  K., Kubiak, M., Szyma{\'n}ski, M.~K., Szewczyk, O., Pietrzy{\'n}ski, G., and
  Soszy{\'n}ski, I.: 2004,
\newblock {\em \apj} {\bf 611}, L33

\bibitem[\protect\astroncite{{Bonazzola} et~al.}{1973}]{bonazzola92}
{Bonazzola}, S., {Perault}, M., {Puget}, J.~L., {Heyvaerts}, J., {Falgarone},
  E., and {Panis}, J.~F.: 1973,
\newblock {\em \jfm} p.~1

\bibitem[\protect\astroncite{{Bonnell} and {Bate}}{2002}]{bonnell02}
{Bonnell}, I.~A. and {Bate}, M.~R.: 2002,
\newblock {\em \mnras} {\bf 336}, 659

\bibitem[\protect\astroncite{{Bonnell} and {Bate}}{2006}]{Bonnell06}
{Bonnell}, I.~A. and {Bate}, M.~R.: 2006,
\newblock {\em \mnras} {\bf 370}, 488

\bibitem[\protect\astroncite{{Bonnell} et~al.}{1997}]{bonnell97}
{Bonnell}, I.~A., {Bate}, M.~R., {Clarke}, C.~J., and {Pringle}, J.~E.: 1997,
\newblock {\em \mnras} {\bf 285}, 201

\bibitem[\protect\astroncite{{Bonnell} et~al.}{2001a}]{bonnell01a}
{Bonnell}, I.~A., {Bate}, M.~R., {Clarke}, C.~J., and {Pringle}, J.~E.: 2001a,
\newblock {\em \mnras} {\bf 323}, 785

\bibitem[\protect\astroncite{{Bonnell} et~al.}{1998}]{bonnell98}
{Bonnell}, I.~A., {Bate}, M.~R., and {Zinnecker}, H.: 1998,
\newblock {\em \mnras} {\bf 298}, 93

\bibitem[\protect\astroncite{Bonnell et~al.}{1998}]{bonbatezin98}
Bonnell, I.~A., Bate, M.~R., and Zinnecker, H.: 1998,
\newblock {\em \mnras} {\bf 298}, 93

\bibitem[\protect\astroncite{{Bonnell} et~al.}{2008}]{bonnell08a}
{Bonnell}, I.~A., {Clark}, P., and {Bate}, M.~R.: 2008,
\newblock {\em \mnras} {\bf 389}, 1556

\bibitem[\protect\astroncite{{Bonnell} et~al.}{2006}]{bonnell06d}
{Bonnell}, I.~A., {Clarke}, C.~J., and {Bate}, M.~R.: 2006,
\newblock {\em \mnras} {\bf 368}, 1296

\bibitem[\protect\astroncite{{Bonnell} et~al.}{2001b}]{bonnell01b}
{Bonnell}, I.~A., {Clarke}, C.~J., {Bate}, M.~R., and {Pringle}, J.~E.: 2001b,
\newblock {\em \mnras} {\bf 324}, 573

\bibitem[\protect\astroncite{{Bonnell} et~al.}{2007}]{bonnell07a}
{Bonnell}, I.~A., {Larson}, R.~B., and {Zinnecker}, H.: 2007,
\newblock in B. {Reipurth}, D. {Jewitt}, and K. {Keil} (eds.), {\em Protostars
  and Planets V, B. Reipurth, D. Jewitt, and K. Keil (eds.), University of
  Arizona Press, Tucson, 951 pp., 2007., p.149-164}, pp 149--164

\bibitem[\protect\astroncite{{Bonnell} et~al.}{2004}]{bonnell04}
{Bonnell}, I.~A., {Vine}, S.~G., and {Bate}, M.~R.: 2004,
\newblock {\em \mnras} {\bf 349}, 735

\bibitem[\protect\astroncite{{Bonnor}}{1956}]{bonnor56}
{Bonnor}, W.~B.: 1956,
\newblock {\em \mnras} {\bf 116}, 351

\bibitem[\protect\astroncite{{Bontemps}}{2001}]{Bontempsetal2001}
{Bontemps}, S., e.: 2001,
\newblock {\em \aap} {\bf 372}, 173

\bibitem[\protect\astroncite{{Bourke} et~al.}{2001}]{bourke01}
{Bourke}, T.~L., {Myers}, P.~C., {Robinson}, G., and {Hyland}, A.~R.: 2001,
\newblock {\em \apj} {\bf 554}, 916

\bibitem[\protect\astroncite{{Brandl} et~al.}{1999}]{brandl99}
{Brandl}, B., {Brandner}, W., {Eisenhauer}, F., {Moffat}, A.~F.~J., {Palla},
  F., and {Zinnecker}, H.: 1999,
\newblock {\em \aap} {\bf 352}, L69

\bibitem[\protect\astroncite{Bromm et~al.}{2009}]{brommetal09}
Bromm, V., Yoshida, N., Hernquist, L., and McKee, C.~F.: 2009,
\newblock {\em \nat} {\bf 459}, 49

\bibitem[\protect\astroncite{{Brunt} et~al.}{2009}]{brunt09}
{Brunt}, C.~M., {Heyer}, M.~H., and {Mac Low}, M.: 2009,
\newblock {\em \aap} {\bf 504}, 883

\bibitem[\protect\astroncite{{Burke} and {Hollenbach}}{1983}]{burke83}
{Burke}, J.~R. and {Hollenbach}, D.~J.: 1983,
\newblock {\em \apj} {\bf 265}, 223

\bibitem[\protect\astroncite{{Burrows} et~al.}{2001}]{burrows01}
{Burrows}, A., {Hubbard}, W.~B., {Lunine}, J.~I., and {Liebert}, J.: 2001,
\newblock {\em Reviews of Modern Physics} {\bf 73}, 719

\bibitem[\protect\astroncite{{Caballero}}{2008}]{caballero08}
{Caballero}, J.~A.: 2008,
\newblock {\em \mnras} {\bf 383}, 750

\bibitem[\protect\astroncite{{Carpenter} et~al.}{1997}]{carpenter97}
{Carpenter}, J.~M., {Meyer}, M.~R., {Dougados}, C., {Strom}, S.~E., and
  {Hillenbrand}, L.~A.: 1997,
\newblock {\em \aj} {\bf 114}, 198

\bibitem[\protect\astroncite{{Cernicharo}}{1991}]{Cernicharo1991}
{Cernicharo}, J.: 1991,
\newblock in {\em NATO ASIC Proc. 342: The Physics of Star Formation and Early
  Stellar Evolution}, p. 287

\bibitem[\protect\astroncite{{Chabrier}}{2003}]{chabrier03}
{Chabrier}, G.: 2003,
\newblock {\em Pub. Ast. Soc. Pacific} {\bf 115}, 763

\bibitem[\protect\astroncite{{Chabrier} and {Hennebelle}}{2010}]{chabrier10a}
{Chabrier}, G. and {Hennebelle}, P.: 2010,
\newblock {\em \apjl} {\bf 725}, L79

\bibitem[\protect\astroncite{{Chandrasekhar}}{1951a}]{1951RSPSA.210...18C}
{Chandrasekhar}, S.: 1951a,
\newblock {\em Royal Society of London Proceedings Series A} {\bf 210}, 18

\bibitem[\protect\astroncite{{Chandrasekhar}}{1951b}]{1951RSPSA.210...26C}
{Chandrasekhar}, S.: 1951b,
\newblock {\em Royal Society of London Proceedings Series A} {\bf 210}, 26

\bibitem[\protect\astroncite{{Chini} et~al.}{2004}]{chini04}
{Chini}, R., {Hoffmeister}, V., {Kimeswenger}, S., {Nielbock}, M.,
  {N{\"u}rnberger}, D., {Schmidtobreick}, L., and {Sterzik}, M.: 2004,
\newblock {\em \nat} {\bf 429}, 155

\bibitem[\protect\astroncite{{Chini} et~al.}{2006}]{chini06}
{Chini}, R., {Hoffmeister}, V.~H., {Nielbock}, M., {Scheyda}, C.~M.,
  {Steinacker}, J., {Siebenmorgen}, R., and {N{\"u}rnberger}, D.: 2006,
\newblock {\em \apjl} {\bf 645}, L61

\bibitem[\protect\astroncite{{Clark} and {Bonnell}}{2006}]{clark06}
{Clark}, P.~C. and {Bonnell}, I.~A.: 2006,
\newblock {\em \mnras} {\bf 368}, 1787

\bibitem[\protect\astroncite{{Clark} et~al.}{2011}]{clark11a}
{Clark}, P.~C., {Glover}, S.~C.~O., {Smith}, R.~J., {Greif}, T.~H., {Klessen},
  R.~S., and {Bromm}, V.: 2011,
\newblock {\em Science} {\bf 331}, 1040

\bibitem[\protect\astroncite{{Clark} et~al.}{2007}]{clark07}
{Clark}, P.~C., {Klessen}, R.~S., and {Bonnell}, I.~A.: 2007,
\newblock {\em \mnras} {\bf 379}, 57

\bibitem[\protect\astroncite{{Crutcher} et~al.}{2003}]{Crutcher03}
{Crutcher}, R., {Heiles}, C., and {Troland}, T.: 2003,
\newblock in E. {Falgarone} and T. {Passot} (eds.), {\em Turbulence and
  Magnetic Fields in Astrophysics}, Vol. 614 of {\em Lecture Notes in Physics,
  Berlin Springer Verlag}, pp 155--181

\bibitem[\protect\astroncite{{Crutcher}}{1999}]{crutcher99}
{Crutcher}, R.~M.: 1999,
\newblock {\em \apj} {\bf 520}, 706

\bibitem[\protect\astroncite{{Crutcher} et~al.}{2009}]{2009ApJ...692..844C}
{Crutcher}, R.~M., {Hakobian}, N., and {Troland}, T.~H.: 2009,
\newblock {\em \apj} {\bf 692}, 844

\bibitem[\protect\astroncite{{Crutcher} et~al.}{2010a}]{2010MNRAS.402L..64C}
{Crutcher}, R.~M., {Hakobian}, N., and {Troland}, T.~H.: 2010a,
\newblock {\em \mnras} {\bf 402}, L64

\bibitem[\protect\astroncite{{Crutcher} and {Troland}}{2000}]{crutcher00}
{Crutcher}, R.~M. and {Troland}, T.~H.: 2000,
\newblock {\em \apjl} {\bf 537}, L139

\bibitem[\protect\astroncite{{Crutcher} et~al.}{1999}]{crutcher99a}
{Crutcher}, R.~M., {Troland}, T.~H., {Lazareff}, B., {Paubert}, G., and
  {Kaz{\`e}s}, I.: 1999,
\newblock {\em \apjl} {\bf 514}, L121

\bibitem[\protect\astroncite{{Crutcher} et~al.}{2010b}]{2010ApJ...725..466C}
{Crutcher}, R.~M., {Wandelt}, B., {Heiles}, C., {Falgarone}, E., and {Troland},
  T.~H.: 2010b,
\newblock {\em \apj} {\bf 725}, 466

\bibitem[\protect\astroncite{Davies et~al.}{2010}]{daviesetal10}
Davies, B., Lumsden, S.~L., Hoare, M.~G., Oudmaijer, R.~D., and de~Wit, W.-J.:
  2010,
\newblock {\em \mnras} {\bf 402}, 1504

\bibitem[\protect\astroncite{{de Avillez} and {Mac Low}}{2002}]{deavillez02}
{de Avillez}, M.~A. and {Mac Low}, M.: 2002,
\newblock {\em \apj} {\bf 581}, 1047

\bibitem[\protect\astroncite{de~Wit et~al.}{2004}]{dewitetal04}
de~Wit, W.~J., Testi, L., Palla, F., Vanzi, L., and Zinnecker, H.: 2004,
\newblock {\em \aap} {\bf 425}, 937

\bibitem[\protect\astroncite{{Di Francesco} et~al.}{2007}]{DiFrancescoetal2007}
{Di Francesco}, J., {Evans}, II, N.~J., {Caselli}, P., {Myers}, P.~C.,
  {Shirley}, Y., {Aikawa}, Y., and {Tafalla}, M.: 2007,
\newblock in B. {Reipurth}, D. {Jewitt}, and K. {Keil} (eds.), {\em Protostars
  and Planets V}, pp 17--32

\bibitem[\protect\astroncite{{Dib} et~al.}{2006}]{dib06}
{Dib}, S., {Bell}, E., and {Burkert}, A.: 2006,
\newblock {\em \apj} {\bf 638}, 797

\bibitem[\protect\astroncite{{Dib} et~al.}{2007}]{Dib2007}
{Dib}, S., {Kim}, J., {V{\'a}zquez-Semadeni}, E., {Burkert}, A., and
  {Shadmehri}, M.: 2007,
\newblock {\em \apj} {\bf 661}, 262

\bibitem[\protect\astroncite{{Dobbs} et~al.}{2005}]{dobbs05}
{Dobbs}, C.~L., {Bonnell}, I.~A., and {Clark}, P.~C.: 2005,
\newblock {\em \mnras} {\bf 360}, 2

\bibitem[\protect\astroncite{{Ebert}}{1955}]{ebert55}
{Ebert}, R.: 1955,
\newblock {\em Zeitschrift fur Astrophysics} {\bf 37}, 217

\bibitem[\protect\astroncite{{Elmegreen}}{2007}]{Elmegreen07}
{Elmegreen}, B.~G.: 2007,
\newblock {\em \apj} {\bf 668}, 1064

\bibitem[\protect\astroncite{{Elmegreen} et~al.}{2008}]{elmegreen08}
{Elmegreen}, B.~G., {Klessen}, R.~S., and {Wilson}, C.~D.: 2008,
\newblock {\em \apj} {\bf 681}, 365

\bibitem[\protect\astroncite{{Elmegreen} and {Scalo}}{2004}]{elmegreen04}
{Elmegreen}, B.~G. and {Scalo}, J.: 2004,
\newblock {\em \araa} {\bf 42}, 211

\bibitem[\protect\astroncite{{Evans}}{1999}]{evans99}
{Evans}, II, N.~J.: 1999,
\newblock {\em \araa} {\bf 37}, 311

\bibitem[\protect\astroncite{{Fatuzzo} and {Adams}}{2002}]{FatuzzoAdams2002}
{Fatuzzo}, M. and {Adams}, F.~C.: 2002,
\newblock {\em \apj} {\bf 570}, 210

\bibitem[\protect\astroncite{{Federrath} et~al.}{2010}]{federrath10}
{Federrath}, C., {Duval}, J., {Klessen}, R., {Schmidt}, W., and {Mac Low},
  M.-M.: 2010,
\newblock {\em Astronomy \& Astrophysics} {\bf 512}, A81

\bibitem[\protect\astroncite{{Federrath} et~al.}{2009}]{federrath09b}
{Federrath}, C., {Klessen}, R.~S., and {Schmidt}, W.: 2009,
\newblock {\em \apj} {\bf 692}, 364

\bibitem[\protect\astroncite{{Federrath} et~al.}{2011}]{2011ApJ...731...62F}
{Federrath}, C., {Sur}, S., {Schleicher}, D.~R.~G., {Banerjee}, R., and
  {Klessen}, R.~S.: 2011,
\newblock {\em \apj} {\bf 731}, 62

\bibitem[\protect\astroncite{{Field} et~al.}{1969}]{Field69}
{Field}, G.~B., {Goldsmith}, D.~W., and {Habing}, H.~J.: 1969,
\newblock {\em \apjl} {\bf 155}, L149

\bibitem[\protect\astroncite{{Figer}}{2005}]{figer05}
{Figer}, D.~F.: 2005,
\newblock {\em \nat} {\bf 434}, 192

\bibitem[\protect\astroncite{{Galv{\'a}n-Madrid}
  et~al.}{2011}]{galvan-madrid11}
{Galv{\'a}n-Madrid}, R., {Peters}, T., {Keto}, E.~R., {Mac Low}, M.-M.,
  {Banerjee}, R., and {Klessen}, R.~S.: 2011,
\newblock {\em \mnras} pp 1081--+

\bibitem[\protect\astroncite{Garay and Lizano}{1999}]{garliz99}
Garay, G. and Lizano, S.: 1999,
\newblock {\em \pasp} {\bf 111}, 1049

\bibitem[\protect\astroncite{{Girichidis} et~al.}{2011}]{girichidis11a}
{Girichidis}, P., {Federrath}, C., {Banerjee}, R., and {Klessen}, R.~S.: 2011,
\newblock {\em \mnras} {\bf 413}, 2741

\bibitem[\protect\astroncite{{Glover}
  et~al.}{2010}]{GloverFederrathMacLowKlessen2010}
{Glover}, S.~C.~O., {Federrath}, C., {Mac Low}, M., and {Klessen}, R.~S.: 2010,
\newblock {\em \mnras} {\bf 404}, 2

\bibitem[\protect\astroncite{{Glover} and {Mac Low}}{2007a}]{Glover07a}
{Glover}, S.~C.~O. and {Mac Low}, M.-M.: 2007a,
\newblock {\em \apjs} {\bf 169}, 239

\bibitem[\protect\astroncite{{Glover} and {Mac Low}}{2007b}]{Glover07b}
{Glover}, S.~C.~O. and {Mac Low}, M.-M.: 2007b,
\newblock {\em \apj} {\bf 659}, 1317

\bibitem[\protect\astroncite{{Goldsmith}}{2001}]{Goldsmith01}
{Goldsmith}, P.~F.: 2001,
\newblock {\em \apj} {\bf 557}, 736

\bibitem[\protect\astroncite{{Goldsmith} and {Langer}}{1978}]{goldsmith78}
{Goldsmith}, P.~F. and {Langer}, W.~D.: 1978,
\newblock {\em \apj} {\bf 222}, 881

\bibitem[\protect\astroncite{{Gomez} et~al.}{1992}]{gomez92}
{Gomez}, M., {Jones}, B.~F., {Hartmann}, L., {Kenyon}, S.~J., {Stauffer},
  J.~R., {Hewett}, R., and {Reid}, I.~N.: 1992,
\newblock {\em \aj} {\bf 104}, 762

\bibitem[\protect\astroncite{{Goodman} et~al.}{1998}]{goodman98}
{Goodman}, A.~A., {Barranco}, J.~A., {Wilner}, D.~J., and {Heyer}, M.~H.: 1998,
\newblock {\em \apj} {\bf 504}, 223

\bibitem[\protect\astroncite{Goto et~al.}{2006}]{gotoetal06}
Goto, M., Stecklum, B., Linz, H., Feldt, M., Henning, T., Pascucci, I., and
  Usuda, T.: 2006,
\newblock {\em \apj} {\bf 649}, 299

\bibitem[\protect\astroncite{{Greene} and {Meyer}}{1995}]{green95}
{Greene}, T.~P. and {Meyer}, M.~R.: 1995,
\newblock {\em \apj} {\bf 450}, 233

\bibitem[\protect\astroncite{{Greif} et~al.}{2011}]{greif11b}
{Greif}, T.~H., {Springel}, V., {White}, S.~D.~M., {Glover}, S.~C.~O., {Clark},
  P.~C., {Smith}, R.~J., {Klessen}, R.~S., and {Bromm}, V.: 2011,
\newblock {\em \apj} {\bf 737}, 75

\bibitem[\protect\astroncite{{Gritschneder} et~al.}{2009}]{gritschneder09a}
{Gritschneder}, M., {Naab}, T., {Burkert}, A., {Walch}, S., {Heitsch}, F., and
  {Wetzstein}, M.: 2009,
\newblock {\em \mnras} {\bf 393}, 21

\bibitem[\protect\astroncite{{Hartmann}}{2002}]{hartmann02}
{Hartmann}, L.: 2002,
\newblock {\em \apj} {\bf 578}, 914

\bibitem[\protect\astroncite{{Hartmann} et~al.}{2001}]{Hartmann01}
{Hartmann}, L., {Ballesteros-Paredes}, J., and {Bergin}, E.~A.: 2001,
\newblock {\em \apj} {\bf 562}, 852

\bibitem[\protect\astroncite{{Heiles} and {Troland}}{2005}]{heiles05}
{Heiles}, C. and {Troland}, T.~H.: 2005,
\newblock {\em \apj} {\bf 624}, 773

\bibitem[\protect\astroncite{{Heitsch} et~al.}{2008}]{2008ApJ...683..786H}
{Heitsch}, F., {Hartmann}, L.~W., and {Burkert}, A.: 2008,
\newblock {\em \apj} {\bf 683}, 786

\bibitem[\protect\astroncite{{Heitsch} et~al.}{2001a}]{Heitschetal2001}
{Heitsch}, F., {Mac Low}, M.-M., and {Klessen}, R.~S.: 2001a,
\newblock {\em \apj} {\bf 547}, 280

\bibitem[\protect\astroncite{{Heitsch} et~al.}{2006a}]{heitsch06a}
{Heitsch}, F., {Slyz}, A.~D., {Devriendt}, J.~E.~G., and {Burkert}, A.: 2006a,
\newblock {\em \mnras} {\bf 373}, 1379

\bibitem[\protect\astroncite{{Heitsch} et~al.}{2006b}]{2006ApJ...648.1052H}
{Heitsch}, F., {Slyz}, A.~D., {Devriendt}, J.~E.~G., {Hartmann}, L.~W., and
  {Burkert}, A.: 2006b,
\newblock {\em \apj} {\bf 648}, 1052

\bibitem[\protect\astroncite{{Heitsch} et~al.}{2001b}]{heitsch01b}
{Heitsch}, F., {Zweibel}, E.~G., {Mac Low}, M.-M., {Li}, P., and {Norman},
  M.~L.: 2001b,
\newblock {\em \apj} {\bf 561}, 800

\bibitem[\protect\astroncite{{Heitsch} et~al.}{2004}]{heitsch04}
{Heitsch}, F., {Zweibel}, E.~G., {Slyz}, A.~D., and {Devriendt}, J.~E.~G.:
  2004,
\newblock {\em \apj} {\bf 603}, 165

\bibitem[\protect\astroncite{{Hennebelle} et~al.}{2008}]{2008A&A...486L..43H}
{Hennebelle}, P., {Banerjee}, R., {V{\'a}zquez-Semadeni}, E., {Klessen}, R.~S.,
  and {Audit}, E.: 2008,
\newblock {\em \aap} {\bf 486}, L43

\bibitem[\protect\astroncite{{Hennebelle} and
  {Chabrier}}{2008}]{HennebelleChabrier2008}
{Hennebelle}, P. and {Chabrier}, G.: 2008,
\newblock {\em \apj} {\bf 684}, 395

\bibitem[\protect\astroncite{{Hennebelle} and {Chabrier}}{2009}]{hennebelle09c}
{Hennebelle}, P. and {Chabrier}, G.: 2009,
\newblock {\em \apj} {\bf 702}, 1428

\bibitem[\protect\astroncite{{Hennebelle} and {Ciardi}}{2009}]{hennebelle09}
{Hennebelle}, P. and {Ciardi}, A.: 2009,
\newblock {\em \aap} {\bf 506}, L29

\bibitem[\protect\astroncite{{Hennebelle} et~al.}{2011}]{hennebelle11}
{Hennebelle}, P., {Commer{\c c}on}, B., {Joos}, M., {Klessen}, R.~S.,
  {Krumholz}, M., {Tan}, J.~C., and {Teyssier}, R.: 2011,
\newblock {\em \aap} {\bf 528}, A72

\bibitem[\protect\astroncite{{Hennebelle} and {Fromang}}{2008}]{hennebelle08a}
{Hennebelle}, P. and {Fromang}, S.: 2008,
\newblock {\em \aap} {\bf 477}, 9

\bibitem[\protect\astroncite{{Hennebelle} and {Teyssier}}{2008}]{hennebelle08c}
{Hennebelle}, P. and {Teyssier}, R.: 2008,
\newblock {\em \aap} {\bf 477}, 25

\bibitem[\protect\astroncite{{Heyer} et~al.}{2008}]{heyer08}
{Heyer}, M., {Krawczyk}, C., {Duval}, J., and {Jackson}, J.~M.: 2008,
\newblock {\em ArXiv e-prints}

\bibitem[\protect\astroncite{{Heyer} and {Brunt}}{2004}]{heyer04a}
{Heyer}, M.~H. and {Brunt}, C.~M.: 2004,
\newblock {\em \apjl} {\bf 615}, L45

\bibitem[\protect\astroncite{{Hillenbrand}}{1997}]{Hillenbrand97}
{Hillenbrand}, L.~A.: 1997,
\newblock {\em \aj} {\bf 113}, 1733

\bibitem[\protect\astroncite{{Hillenbrand} and
  {Hartmann}}{1998}]{hillenbrand98}
{Hillenbrand}, L.~A. and {Hartmann}, L.~W.: 1998,
\newblock {\em \apj} {\bf 492}, 540

\bibitem[\protect\astroncite{{Hirota} et~al.}{2007}]{hirota07}
{Hirota}, T., {Bushimata}, T., {Choi}, Y.~K., {Honma}, M., {Imai}, H.,
  {Iwadate}, K., {Jike}, T., {Kameno}, S., {Kameya}, O., {Kamohara}, R.,
  {Kan-Ya}, Y., {Kawaguchi}, N., {Kijima}, M., {Kim}, M.~K., {Kobayashi}, H.,
  {Kuji}, S., {Kurayama}, T., {Manabe}, S., {Maruyama}, K., {Matsui}, M.,
  {Matsumoto}, N., and {Miyaji}, T.: 2007,
\newblock {\em \pasj} {\bf 59}, 897

\bibitem[\protect\astroncite{Ho and Haschick}{1981}]{hohaschik81}
Ho, P.~T.~P. and Haschick, A.~D.: 1981,
\newblock {\em \apj} {\bf 248}, 622

\bibitem[\protect\astroncite{{Hunter} et~al.}{1995}]{hunter95}
{Hunter}, D.~A., {Shaya}, E.~J., {Scowen}, P., {Hester}, J.~J., {Groth}, E.~J.,
  {Lynds}, R., and {O'Neil}, Jr., E.~J.: 1995,
\newblock {\em \apj} {\bf 444}, 758

\bibitem[\protect\astroncite{{Jappsen} et~al.}{2005}]{Jappsen05}
{Jappsen}, A.-K., {Klessen}, R.~S., {Larson}, R.~B., {Li}, Y., and {Mac Low},
  M.-M.: 2005,
\newblock {\em \aap} {\bf 435}, 611

\bibitem[\protect\astroncite{Jiang et~al.}{2008}]{jiangetal08}
Jiang, Z., Tamura, M., Hoare, M.~G., Yao, Y., Ishii, M., Fang, M., and Yang,
  J.: 2008,
\newblock {\em \apj} {\bf 673}, L175

\bibitem[\protect\astroncite{{Jijina} et~al.}{1999}]{jijina99}
{Jijina}, J., {Myers}, P.~C., and {Adams}, F.~C.: 1999,
\newblock {\em \apjs} {\bf 125}, 161

\bibitem[\protect\astroncite{{Johnstone} et~al.}{2001}]{johnstone01}
{Johnstone}, D., {Fich}, M., {Mitchell}, G.~F., and {Moriarty-Schieven}, G.:
  2001,
\newblock {\em \apj} {\bf 559}, 307

\bibitem[\protect\astroncite{{Johnstone} et~al.}{2006}]{Johnstoneetal2006}
{Johnstone}, D., {Matthews}, H., and {Mitchell}, G.~F.: 2006,
\newblock {\em \apj} {\bf 639}, 259

\bibitem[\protect\astroncite{{Johnstone} et~al.}{2000}]{johnstone00}
{Johnstone}, D., {Wilson}, C.~D., {Moriarty-Schieven}, G., {Joncas}, G.,
  {Smith}, G., {Gregersen}, E., and {Fich}, M.: 2000,
\newblock {\em \apj} {\bf 545}, 327

\bibitem[\protect\astroncite{{Kahn}}{1974}]{kahn74}
{Kahn}, F.~D.: 1974,
\newblock {\em \aap} {\bf 37}, 149

\bibitem[\protect\astroncite{{Kandori} et~al.}{2005}]{kandori05}
{Kandori}, R., {Nakajima}, Y., {Tamura}, M., {Tatematsu}, K., {Aikawa}, Y.,
  {Naoi}, T., {Sugitani}, K., {Nakaya}, H., {Nagayama}, T., {Nagata}, T.,
  {Kurita}, M., {Kato}, D., {Nagashima}, C., and {Sato}, S.: 2005,
\newblock {\em \aj} {\bf 130}, 2166

\bibitem[\protect\astroncite{Keto}{2002}]{keto02b}
Keto, E.: 2002,
\newblock {\em \apj} {\bf 580}, 980

\bibitem[\protect\astroncite{{Keto}}{2003}]{keto03}
{Keto}, E.: 2003,
\newblock {\em \apj} {\bf 599}, 1196

\bibitem[\protect\astroncite{{Keto}}{2007}]{keto07}
{Keto}, E.: 2007,
\newblock {\em \apj} {\bf 666}, 976

\bibitem[\protect\astroncite{{Keto} and {Caselli}}{2008}]{keto08}
{Keto}, E. and {Caselli}, P.: 2008,
\newblock {\em ArXiv e-prints} 804

\bibitem[\protect\astroncite{{Keto} and {Field}}{2005}]{keto05}
{Keto}, E. and {Field}, G.: 2005,
\newblock {\em \apj} {\bf 635}, 1151

\bibitem[\protect\astroncite{Keto and Wood}{2006}]{ketoetal06}
Keto, E. and Wood, K.: 2006,
\newblock {\em \apj} {\bf 637}, 850

\bibitem[\protect\astroncite{{Kim} and {Ostriker}}{2001}]{kim01}
{Kim}, W. and {Ostriker}, E.~C.: 2001,
\newblock {\em \apj} {\bf 559}, 70

\bibitem[\protect\astroncite{{Kippenhahn} and {Weigert}}{1994}]{kippenhahn94}
{Kippenhahn}, R. and {Weigert}, A.: 1994,
\newblock {\em Stellar Structure and Evolution},
\newblock Springer-Verlag, Berlin

\bibitem[\protect\astroncite{{Kirk} et~al.}{2007}]{kirk07}
{Kirk}, H., {Johnstone}, D., and {Tafalla}, M.: 2007,
\newblock {\em \apj} {\bf 668}, 1042

\bibitem[\protect\astroncite{Klessen}{2001}]{klessen01}
Klessen, R.~S.: 2001,
\newblock {\em \apj} {\bf 556}, 837

\bibitem[\protect\astroncite{{Klessen}}{2001a}]{Klessen2001}
{Klessen}, R.~S.: 2001a,
\newblock {\em \apj} {\bf 556}, 837

\bibitem[\protect\astroncite{{Klessen}}{2001b}]{Klessen2001b}
{Klessen}, R.~S.: 2001b,
\newblock {\em \apjl} {\bf 550}, L77

\bibitem[\protect\astroncite{{Klessen} et~al.}{2005}]{klessen05}
{Klessen}, R.~S., {Ballesteros-Paredes}, J., {V{\'a}zquez-Semadeni}, E., and
  {Dur{\'a}n-Rojas}, C.: 2005,
\newblock {\em \apj} {\bf 620}, 786

\bibitem[\protect\astroncite{{Klessen} and {Burkert}}{2000}]{klessen00a}
{Klessen}, R.~S. and {Burkert}, A.: 2000,
\newblock {\em \apjs} {\bf 128}, 287

\bibitem[\protect\astroncite{{Klessen} and {Burkert}}{2001}]{klessen01s}
{Klessen}, R.~S. and {Burkert}, A.: 2001,
\newblock {\em \apj} {\bf 549}, 386

\bibitem[\protect\astroncite{{Klessen} et~al.}{2000}]{klessen00b}
{Klessen}, R.~S., {Heitsch}, F., and {Mac Low}, M.: 2000,
\newblock {\em \apj} {\bf 535}, 887

\bibitem[\protect\astroncite{{Klessen} and {Hennebelle}}{2010}]{klessen10}
{Klessen}, R.~S. and {Hennebelle}, P.: 2010,
\newblock {\em \aap} {\bf 520}, A17

\bibitem[\protect\astroncite{{Klessen} et~al.}{2011}]{klessen11}
{Klessen}, R.~S., {Krumholz}, M.~R., and {Heitsch}, F.: 2011,
\newblock {\em \asl} {\bf 4}, 258

\bibitem[\protect\astroncite{{Klessen} and {Lin}}{2003}]{klessen03b}
{Klessen}, R.~S. and {Lin}, D.~N.: 2003,
\newblock {\em \pre} {\bf 67}, 046311

\bibitem[\protect\astroncite{{Kn\"odlseder}}{2000}]{knoedlseder00}
{Kn\"odlseder}, J.: 2000,
\newblock {\em \aap} {\bf 360}, 539

\bibitem[\protect\astroncite{{Kolmogorov}}{1941}]{Kolmogorov1941}
{Kolmogorov}, A.~N.: 1941,
\newblock {\em Dokl. Akad. Nauk SSSR} {\bf 30}, 301

\bibitem[\protect\astroncite{{Kramer} et~al.}{1998}]{kramer98a}
{Kramer}, C., {Stutzki}, J., {Rohrig}, R., and {Corneliussen}, U.: 1998,
\newblock {\em \aap} {\bf 329}, 249

\bibitem[\protect\astroncite{{Kroupa}}{2002}]{kroupa02}
{Kroupa}, P.: 2002,
\newblock {\em Science} {\bf 295}, 82

\bibitem[\protect\astroncite{{Krumholz}}{2006}]{krumholz06b}
{Krumholz}, M.~R.: 2006,
\newblock {\em \apjl} {\bf 641}, L45

\bibitem[\protect\astroncite{{Krumholz} et~al.}{2007a}]{krumholz07a}
{Krumholz}, M.~R., {Klein}, R.~I., and {McKee}, C.~F.: 2007a,
\newblock {\em \apj} {\bf 656}, 959

\bibitem[\protect\astroncite{Krumholz et~al.}{2009}]{krumholzetal09}
Krumholz, M.~R., Klein, R.~I., McKee, C.~F., Offner, S.~S.~R., and Cunningham,
  A.~J.: 2009,
\newblock {\em {Science}} {\bf 323}, 754

\bibitem[\protect\astroncite{{Krumholz} et~al.}{2006}]{krumholz06d}
{Krumholz}, M.~R., {Matzner}, C.~D., and {McKee}, C.~F.: 2006,
\newblock {\em \apj} {\bf 653}, 361

\bibitem[\protect\astroncite{{Krumholz} and {McKee}}{2005}]{krumholz05c}
{Krumholz}, M.~R. and {McKee}, C.~F.: 2005,
\newblock {\em \apj} {\bf 630}, 250

\bibitem[\protect\astroncite{{Krumholz} and {McKee}}{2008}]{krumholz08a}
{Krumholz}, M.~R. and {McKee}, C.~F.: 2008,
\newblock {\em \nat}

\bibitem[\protect\astroncite{{Krumholz} et~al.}{2007b}]{krumholz07f}
{Krumholz}, M.~R., {Stone}, J.~M., and {Gardiner}, T.~A.: 2007b,
\newblock {\em \apj} {\bf 671}, 518

\bibitem[\protect\astroncite{{Kuiper} et~al.}{2010}]{kuiper10}
{Kuiper}, R., {Klahr}, H., {Beuther}, H., and {Henning}, T.: 2010,
\newblock {\em \apj} {\bf 722}, 1556

\bibitem[\protect\astroncite{{Kuiper} et~al.}{2011}]{kuiper11}
{Kuiper}, R., {Klahr}, H., {Beuther}, H., and {Henning}, T.: 2011,
\newblock {\em ArXiv e-prints}

\bibitem[\protect\astroncite{Kurtz et~al.}{2000}]{kurtzetal00}
Kurtz, S., Cesaroni, R., Churchwell, E., Hofner, P., and Walmsley, C.~M.: 2000,
\newblock in V. Mannings, A.~P. Boss, and S.~S. Russell (eds.), {\em
  {Protostars and Planets IV}}, pp 299--326, {The University of Arizona Press}

\bibitem[\protect\astroncite{Kurtz et~al.}{1994}]{kurtzetal94}
Kurtz, S., Churchwell, E., and Wood, D.~O.~S.: 1994,
\newblock {\em \apjs} {\bf 91}, 659

\bibitem[\protect\astroncite{{Lada}}{2006}]{lada06}
{Lada}, C.~J.: 2006,
\newblock {\em \apjl} {\bf 640}, L63

\bibitem[\protect\astroncite{{Lada} et~al.}{2003}]{lada03b}
{Lada}, C.~J., {Bergin}, E.~A., {Alves}, J.~F., and {Huard}, T.~L.: 2003,
\newblock {\em \apj} {\bf 586}, 286

\bibitem[\protect\astroncite{{Lada} and {Lada}}{2003}]{lada03}
{Lada}, C.~J. and {Lada}, E.~A.: 2003,
\newblock {\em \araa} {\bf 41}, 57

\bibitem[\protect\astroncite{Lada and Lada}{2003}]{ladalada03}
Lada, C.~J. and Lada, E.~A.: 2003,
\newblock {\em \araa} {\bf 41}, 57

\bibitem[\protect\astroncite{{Lada} et~al.}{2008}]{lada08a}
{Lada}, C.~J., {Muench}, A.~A., {Rathborne}, J., {Alves}, J.~F., and
  {Lombardi}, M.: 2008,
\newblock {\em \apj} {\bf 672}, 410

\bibitem[\protect\astroncite{{Lada} et~al.}{2006}]{Ladaetal2006}
{Lada}, C.~L., {Alves}, J.~F., {Lombardi}, M., and {Lada}, E.~A.: 2006,
\newblock in {\em Protostars and Planets V, {\it{in press}}}

\bibitem[\protect\astroncite{{Lai} et~al.}{2001}]{lai01}
{Lai}, S., {Crutcher}, R.~M., {Girart}, J.~M., and {Rao}, R.: 2001,
\newblock {\em \apj} {\bf 561}, 864

\bibitem[\protect\astroncite{{Lai} et~al.}{2002}]{lai02}
{Lai}, S., {Crutcher}, R.~M., {Girart}, J.~M., and {Rao}, R.: 2002,
\newblock {\em \apj} {\bf 566}, 925

\bibitem[\protect\astroncite{{Langer} et~al.}{2005}]{langer05}
{Langer}, W.~D., {Velusamy}, T., {Li}, D., and {Goldsmith}, P.~F.: 2005,
\newblock in {\em Protostars and Planets V}, p. 8179

\bibitem[\protect\astroncite{{Larson}}{1981}]{larson81}
{Larson}, R.~B.: 1981,
\newblock {\em \mnras} {\bf 194}, 809

\bibitem[\protect\astroncite{{Larson}}{2005}]{Larson05}
{Larson}, R.~B.: 2005,
\newblock {\em \mnras} {\bf 359}, 211

\bibitem[\protect\astroncite{{Larson}}{2007}]{Larson07}
{Larson}, R.~B.: 2007,
\newblock {\em Reports on Progress in Physics} {\bf 70}, 337

\bibitem[\protect\astroncite{Larson and Starrfield}{1971}]{larsstarr71}
Larson, R.~B. and Starrfield, S.: 1971,
\newblock {\em \aap} {\bf 13}, 190

\bibitem[\protect\astroncite{{Lee} et~al.}{1999}]{lee99}
{Lee}, C.~W., {Myers}, P.~C., and {Tafalla}, M.: 1999,
\newblock {\em \apj} {\bf 526}, 788

\bibitem[\protect\astroncite{{Lesieur}}{1997}]{lesieur97}
{Lesieur}, M.: 1997,
\newblock {\em {Turbulence in Fluids}},
\newblock Kluwer Academic Publishers, Dordrecht

\bibitem[\protect\astroncite{{Li} et~al.}{2003}]{li03}
{Li}, Y., {Klessen}, R.~S., and {Mac Low}, M.-M.: 2003,
\newblock {\em \apj} {\bf 592}, 975

\bibitem[\protect\astroncite{{Li} and {Nakamura}}{2004}]{zli04}
{Li}, Z. and {Nakamura}, F.: 2004,
\newblock {\em \apjl} {\bf 609}, L83

\bibitem[\protect\astroncite{{Li} and {Nakamura}}{2006}]{li06b}
{Li}, Z.-Y. and {Nakamura}, F.: 2006,
\newblock {\em \apjl} {\bf 640}, L187

\bibitem[\protect\astroncite{Linz et~al.}{2005}]{linzetal05}
Linz, H., Stecklum, B., Henning, T., Hofner, P., and Brandl, B.: 2005,
\newblock {\em \aap} {\bf 429}, 903

\bibitem[\protect\astroncite{{Mac Low} and {Klessen}}{2004}]{maclow04}
{Mac Low}, M. and {Klessen}, R.~S.: 2004,
\newblock {\em Reviews of Modern Physics} {\bf 76}, 125

\bibitem[\protect\astroncite{{Machida}}{2008}]{machida08}
{Machida}, M.~N.: 2008,
\newblock {\em \apjl} {\bf 682}, L1

\bibitem[\protect\astroncite{{Maret} et~al.}{2007}]{maret07a}
{Maret}, S., {Bergin}, E.~A., and {Lada}, C.~J.: 2007,
\newblock {\em \apjl} {\bf 670}, L25

\bibitem[\protect\astroncite{{Massey}}{2003}]{massey03}
{Massey}, P.: 2003,
\newblock {\em \araa} {\bf 41}, 15

\bibitem[\protect\astroncite{{Matzner}}{2002}]{matzner02}
{Matzner}, C.~D.: 2002,
\newblock {\em \apj} {\bf 566}, 302

\bibitem[\protect\astroncite{{Matzner} and {McKee}}{2000}]{matzner00}
{Matzner}, C.~D. and {McKee}, C.~F.: 2000,
\newblock {\em \apj} {\bf 545}, 364

\bibitem[\protect\astroncite{{McCaughrean}}{2001}]{mccaughrean01}
{McCaughrean}, M.: 2001,
\newblock in T. {Montmerle} and P. {Andr{\'{e}}} (eds.), {\em From Darkness to
  Light: Origin and Evolution of Young Stellar Clusters}, Vol. 243 of {\em
  Astronomical Society of the Pacific Conference Series}, p. 449

\bibitem[\protect\astroncite{{McKee}}{1999}]{mckee99}
{McKee}, C.~F.: 1999,
\newblock in C.~J. {Lada} and N.~D. {Kylafis} (eds.), {\em NATO ASIC Proc. 540:
  The Origin of Stars and Planetary Systems}, p.~29

\bibitem[\protect\astroncite{{McKee} and {Ostriker}}{2007}]{mckee07}
{McKee}, C.~F. and {Ostriker}, E.~C.: 2007,
\newblock {\em \araa} {\bf 45}, 565

\bibitem[\protect\astroncite{{McKee} and {Ostriker}}{1977}]{mckee77}
{McKee}, C.~F. and {Ostriker}, J.~P.: 1977,
\newblock {\em \apj} {\bf 218}, 148

\bibitem[\protect\astroncite{{Mellon} and {Li}}{2008}]{mellon08b}
{Mellon}, R.~R. and {Li}, Z.-Y.: 2008,
\newblock {\em ArXiv e-prints}

\bibitem[\protect\astroncite{{Menten} et~al.}{2007}]{menten07}
{Menten}, K.~M., {Reid}, M.~J., {Forbrich}, J., and {Brunthaler}, A.: 2007,
\newblock {\em \aap} {\bf 474}, 515

\bibitem[\protect\astroncite{{Miller} and {Scalo}}{1979}]{miller79}
{Miller}, G.~E. and {Scalo}, J.: 1979,
\newblock {\em \apjs} {\bf 41}, 513

\bibitem[\protect\astroncite{{Moffat} et~al.}{2002}]{moffat02}
{Moffat}, A.~F.~J., {Corcoran}, M.~F., {Stevens}, I.~R., {Skalkowski}, G.,
  {Marchenko}, S.~V., {M{\"u}cke}, A., {Ptak}, A., {Koribalski}, B.~S.,
  {Brenneman}, L., {Mushotzky}, R., {Pittard}, J.~M., {Pollock}, A.~M.~T., and
  {Brandner}, W.: 2002,
\newblock {\em \apj} {\bf 573}, 191

\bibitem[\protect\astroncite{{Morata} et~al.}{2005}]{Morata2005}
{Morata}, O., {Girart}, J.~M., and {Estalella}, R.: 2005,
\newblock {\em \aap} {\bf 435}, 113

\bibitem[\protect\astroncite{{Motte} et~al.}{1998}]{motte98}
{Motte}, F., {Andre}, P., and {Neri}, R.: 1998,
\newblock {\em \aap} {\bf 336}, 150

\bibitem[\protect\astroncite{Motte et~al.}{2008}]{motteetal08}
Motte, F., Bontemps, S., Schneider, N., Schilke, P., and Menten, K.~M.: 2008,
\newblock in H. Beuther, H. Linz, and T. Henning (eds.), {\em {Massive Star
  Formation: Observations Confront Theory}}, Vol. 387 of {\em {Astronomical
  Society of the Pacific Conference Series}}, pp 22--29

\bibitem[\protect\astroncite{{Mouschovias}}{1976}]{mouschovias76}
{Mouschovias}, T.~C.: 1976,
\newblock {\em \apj} {\bf 207}, 141

\bibitem[\protect\astroncite{{Mouschovias}}{1979}]{mouschovias79}
{Mouschovias}, T.~C.: 1979,
\newblock {\em \apj} {\bf 228}, 475

\bibitem[\protect\astroncite{{Mouschovias}}{1991a}]{mouschovias91b}
{Mouschovias}, T.~C.: 1991a,
\newblock in C.~J. {Lada} and N.~D. {Kylafis} (eds.), {\em NATO ASIC Proc. 342:
  The Physics of Star Formation and Early Stellar Evolution}, p.~61

\bibitem[\protect\astroncite{{Mouschovias}}{1991b}]{mouschovias91a}
{Mouschovias}, T.~C.: 1991b,
\newblock in C.~J. {Lada} and N.~D. {Kylafis} (eds.), {\em NATO ASIC Proc. 342:
  The Physics of Star Formation and Early Stellar Evolution}, p. 449

\bibitem[\protect\astroncite{{Mouschovias} and
  {Paleologou}}{1981}]{mouschovias81}
{Mouschovias}, T.~C. and {Paleologou}, E.~V.: 1981,
\newblock {\em \apj} {\bf 246}, 48

\bibitem[\protect\astroncite{{Mouschovias} and
  {Spitzer}}{1976}]{1976ApJ...210..326M}
{Mouschovias}, T.~C. and {Spitzer}, Jr., L.: 1976,
\newblock {\em \apj} {\bf 210}, 326

\bibitem[\protect\astroncite{{Myers}}{1983}]{myers83}
{Myers}, P.~C.: 1983,
\newblock {\em \apj} {\bf 270}, 105

\bibitem[\protect\astroncite{{Nakamura} and {Li}}{2007}]{Nakamura07}
{Nakamura}, F. and {Li}, Z.-Y.: 2007,
\newblock {\em \apj} {\bf 662}, 395

\bibitem[\protect\astroncite{{Nutter} and
  {Ward-Thompson}}{2007}]{NutterWardThompson2007}
{Nutter}, D. and {Ward-Thompson}, D.: 2007,
\newblock {\em \mnras} {\bf 374}, 1413

\bibitem[\protect\astroncite{{Oey} and {Clarke}}{2005}]{oey05}
{Oey}, M.~S. and {Clarke}, C.~J.: 2005,
\newblock {\em \apjl} {\bf 620}, L43

\bibitem[\protect\astroncite{{Offner} et~al.}{2008}]{offner08b}
{Offner}, S.~S.~R., {Klein}, R.~I., and {McKee}, C.~F.: 2008,
\newblock {\em \apj}

\bibitem[\protect\astroncite{{Ossenkopf} and {Mac Low}}{2002}]{Ossenkopf02}
{Ossenkopf}, V. and {Mac Low}, M.-M.: 2002,
\newblock {\em \aap} {\bf 390}, 307

\bibitem[\protect\astroncite{{Padoan}
  et~al.}{2004}]{PadoanJimenezNordlundBoldyrev2004}
{Padoan}, P., {Jimenez}, R., {Nordlund}, {\AA}., and {Boldyrev}, S.: 2004,
\newblock {\em \prl} {\bf 92}, 191102

\bibitem[\protect\astroncite{{Padoan} et~al.}{2001}]{padoan01}
{Padoan}, P., {Juvela}, M., {Goodman}, A.~A., and {Nordlund}, {\AA}.: 2001,
\newblock {\em \apj} {\bf 553}, 227

\bibitem[\protect\astroncite{{Padoan} and {Nordlund}}{1999}]{padoan99}
{Padoan}, P. and {Nordlund}, {\AA}.: 1999,
\newblock {\em \apj} {\bf 526}, 279

\bibitem[\protect\astroncite{{Padoan} and {Nordlund}}{2002}]{Padoan02}
{Padoan}, P. and {Nordlund}, {\AA}.: 2002,
\newblock {\em \apj} {\bf 576}, 870

\bibitem[\protect\astroncite{{Padoan} et~al.}{2007}]{Padoan07}
{Padoan}, P., {Nordlund}, {\AA}., {Kritsuk}, A.~G., {Norman}, M.~L., and {Li},
  P.~S.: 2007,
\newblock {\em \apj} {\bf 661}, 972

\bibitem[\protect\astroncite{{Palla} and {Stahler}}{1999}]{palla99}
{Palla}, F. and {Stahler}, S.~W.: 1999,
\newblock {\em \apj} {\bf 525}, 772

\bibitem[\protect\astroncite{{Panis} and {P\'erault}}{1998}]{panis98}
{Panis}, J.-F. and {P\'erault}, M.: 1998,
\newblock {\em \pfl} {\bf 10}, 3111

\bibitem[\protect\astroncite{{Pavlyuchenkov} et~al.}{2006}]{pavlyuchenkov06}
{Pavlyuchenkov}, Y., {Wiebe}, D., {Launhardt}, R., and {Henning}, T.: 2006,
\newblock {\em \apj} {\bf 645}, 1212

\bibitem[\protect\astroncite{{Persi} et~al.}{2000}]{persi00}
{Persi}, P., {Marenzi}, A.~R., {Olofsson}, G., {Kaas}, A.~A., {Nordh}, L.,
  {Huldtgren}, M., {Abergel}, A., {Andr{\'e}}, P., {Bontemps}, S., {Boulanger},
  F., {Burggdorf}, M., {Casali}, M.~M., {Cesarsky}, C.~J., {Copet}, E.,
  {Davies}, J., {Falgarone}, E., {Montmerle}, T., {Perault}, M., {Prusti}, T.,
  {Puget}, J.~L., and {Sibille}, F.: 2000,
\newblock {\em \aap} {\bf 357}, 219

\bibitem[\protect\astroncite{{Peters} et~al.}{2008}]{peters08a}
{Peters}, T., {Banerjee}, R., and {Klessen}, R.~S.: 2008,
\newblock {\em Physica Scripta Volume T} {\bf 132}, 014026

\bibitem[\protect\astroncite{{Peters} et~al.}{2011}]{Peters2011}
{Peters}, T., {Banerjee}, R., {Klessen}, R.~S., and {Mac Low}, M.: 2011,
\newblock {\em \apj} {\bf 729}, 72

\bibitem[\protect\astroncite{{Peters} et~al.}{2010a}]{Peters2010a}
{Peters}, T., {Banerjee}, R., {Klessen}, R.~S., {Mac Low}, M.-M.,
  {Galv{\'a}n-Madrid}, R., and {Keto}, E.~R.: 2010a,
\newblock {\em ApJ} {\bf 711}, 1017

\bibitem[\protect\astroncite{{Peters} et~al.}{2010b}]{Peters2010c}
{Peters}, T., {Klessen}, R.~S., {Mac Low}, M., and {Banerjee}, R.: 2010b,
\newblock {\em \apj} {\bf 725}, 134

\bibitem[\protect\astroncite{{Peters} et~al.}{2010c}]{Peters2010b}
{Peters}, T., {Mac Low}, M.-M., {Banerjee}, R., {Klessen}, R.~S., and
  {Dullemond}, C.~P.: 2010c,
\newblock {\em ApJ} {\bf 719}, 831

\bibitem[\protect\astroncite{{Poppel}}{1997}]{poeppel97}
{Poppel}, W.: 1997,
\newblock {\em Fundamentals of Cosmic Physics} {\bf 18}, 1

\bibitem[\protect\astroncite{Portegies~Zwart et~al.}{2010}]{pozwetal10}
Portegies~Zwart, S.~F., McMillan, S.~L.~W., and Gieles, M.: 2010,
\newblock {\em \araa} {\bf 48}, 431

\bibitem[\protect\astroncite{{Price} and {Bate}}{2007a}]{price07b}
{Price}, D.~J. and {Bate}, M.~R.: 2007a,
\newblock {\em \apss} {\bf 311}, 75

\bibitem[\protect\astroncite{{Price} and {Bate}}{2007b}]{price07a}
{Price}, D.~J. and {Bate}, M.~R.: 2007b,
\newblock {\em \mnras} {\bf 377}, 77

\bibitem[\protect\astroncite{{Price} and {Bate}}{2008}]{price08a}
{Price}, D.~J. and {Bate}, M.~R.: 2008,
\newblock {\em \mnras} {\bf 385}, 1820

\bibitem[\protect\astroncite{Rauw et~al.}{2005}]{rauwetal05}
Rauw, G., Crowther, P.~A., De~Becker, M., Gosset, E., Naz{\'e}, Y., Sana, H.,
  van~der Hucht, K.~A., Vreux, J.-M., and Williams, P.~M.: 2005,
\newblock {\em \aap} {\bf 432}, 985

\bibitem[\protect\astroncite{{Rosolowsky} et~al.}{2008}]{rosolowsky08a}
{Rosolowsky}, E.~W., {Pineda}, J.~E., {Foster}, J.~B., {Borkin}, M.~A.,
  {Kauffmann}, J., {Caselli}, P., {Myers}, P.~C., and {Goodman}, A.~A.: 2008,
\newblock {\em \apjs} {\bf 175}, 509

\bibitem[\protect\astroncite{{Salpeter}}{1955}]{Salpeter1955}
{Salpeter}, E.~E.: 1955,
\newblock {\em \apj} {\bf 121}, 161

\bibitem[\protect\astroncite{{Sandstrom} et~al.}{2007}]{sandstrom07}
{Sandstrom}, K.~M., {Peek}, J.~E.~G., {Bower}, G.~C., {Bolatto}, A.~D., and
  {Plambeck}, R.~L.: 2007,
\newblock {\em \apj} {\bf 667}, 1161

\bibitem[\protect\astroncite{{Scalo}}{1986}]{scalo86}
{Scalo}, J.: 1986,
\newblock {\em {Fund. of Cosmic Physics}} {\bf 11}, 1

\bibitem[\protect\astroncite{{Scalo} and {Elmegreen}}{2004}]{Scalo04}
{Scalo}, J. and {Elmegreen}, B.~G.: 2004,
\newblock {\em \araa} {\bf 42}, 275

\bibitem[\protect\astroncite{{Scalo} et~al.}{1998}]{scalo98b}
{Scalo}, J., {Vazquez-Semadeni}, E., {Chappell}, D., and {Passot}, T.: 1998,
\newblock {\em \apj} {\bf 504}, 835

\bibitem[\protect\astroncite{{Schleicher} et~al.}{2010}]{2010A&A...522A.115S}
{Schleicher}, D.~R.~G., {Banerjee}, R., {Sur}, S., {Arshakian}, T.~G.,
  {Klessen}, R.~S., {Beck}, R., and {Spaans}, M.: 2010,
\newblock {\em \aap} {\bf 522}, A115

\bibitem[\protect\astroncite{{Schmeja} and
  {Klessen}}{2004}]{SchmejaKlessen2004}
{Schmeja}, S. and {Klessen}, R.~S.: 2004,
\newblock {\em \aap} {\bf 419}, 405

\bibitem[\protect\astroncite{{Schmidt} et~al.}{2008}]{schmidt08}
{Schmidt}, W., {Federrath}, C., and {Klessen}, R.: 2008,
\newblock {\em \prl} {\bf 101}, 194505

\bibitem[\protect\astroncite{{Schneider} et~al.}{2006}]{schneider06}
{Schneider}, N., {Bontemps}, S., {Simon}, R., {Jakob}, H., {Motte}, F.,
  {Miller}, M., {Kramer}, C., and {Stutzki}, J.: 2006,
\newblock {\em \aap} {\bf 458}, 588

\bibitem[\protect\astroncite{{Schneider} et~al.}{2011}]{schneider11}
{Schneider}, N., {Bontemps}, S., {Simon}, R., {Ossenkopf}, V., {Federrath}, C.,
  {Klessen}, R.~S., {Motte}, F., {Andr{\'e}}, P., {Stutzki}, J., and {Brunt},
  C.: 2011,
\newblock {\em \aap} {\bf 529}, A1

\bibitem[\protect\astroncite{{Schneider} et~al.}{2007}]{schneider07}
{Schneider}, N., {Simon}, R., {Bontemps}, S., {Comer\'on}, F., and {Motte}, F.:
  2007,
\newblock {\em \aap} {\bf 474}, 873

\bibitem[\protect\astroncite{Selman and Melnick}{2008}]{selmel08}
Selman, F.~J. and Melnick, J.: 2008,
\newblock {\em \apj} {\bf 689}, 816

\bibitem[\protect\astroncite{{She} and {Leveque}}{1994}]{SheLeveque1994}
{She}, Z.-S. and {Leveque}, E.: 1994,
\newblock {\em \prl} {\bf 72}, 336

\bibitem[\protect\astroncite{{Shu} et~al.}{1987}]{shu87}
{Shu}, F.~H., {Adams}, F.~C., and {Lizano}, S.: 1987,
\newblock {\em \araa} {\bf 25}, 23

\bibitem[\protect\astroncite{{Smith} et~al.}{2008}]{smith08a}
{Smith}, R.~J., {Clark}, P.~C., and {Bonnell}, I.~A.: 2008,
\newblock {\em MNRAS}

\bibitem[\protect\astroncite{{Solomon} et~al.}{1987}]{solomon87}
{Solomon}, P.~M., {Rivolo}, A.~R., {Barrett}, J., and {Yahil}, A.: 1987,
\newblock {\em \apj} {\bf 319}, 730

\bibitem[\protect\astroncite{{Spaans} and {Silk}}{2000}]{spaans00}
{Spaans}, M. and {Silk}, J.: 2000,
\newblock {\em \apj} {\bf 538}, 115

\bibitem[\protect\astroncite{{Spitzer}}{1978}]{spitzer78}
{Spitzer}, L.: 1978,
\newblock {\em Physical Processes in the Interstellar Medium},
\newblock Wiley-Interscience, New York

\bibitem[\protect\astroncite{{Stamatellos} et~al.}{2007}]{stamatellos07a}
{Stamatellos}, D., {Whitworth}, A.~P., and {Ward-Thompson}, D.: 2007,
\newblock {\em \mnras} {\bf 379}, 1390

\bibitem[\protect\astroncite{{Stutzki} and
  {G{\"{u}}sten}}{1990}]{StutzkiGuesten1990}
{Stutzki}, J. and {G{\"{u}}sten}, R.: 1990,
\newblock {\em \apj} {\bf 356}, 513

\bibitem[\protect\astroncite{{Sugitani} et~al.}{2010}]{sugitani10}
{Sugitani}, K., {Nakamura}, F., {Tamura}, M., {Watanabe}, M., {Kandori}, R.,
  {Nishiyama}, S., {Kusakabe}, N., {Hashimoto}, J., {Nagata}, T., and {Sato},
  S.: 2010,
\newblock {\em \apj} {\bf 716}, 299

\bibitem[\protect\astroncite{{Sur} et~al.}{2010}]{2010ApJ...721L.134S}
{Sur}, S., {Schleicher}, D.~R.~G., {Banerjee}, R., {Federrath}, C., and
  {Klessen}, R.~S.: 2010,
\newblock {\em \apjl} {\bf 721}, L134

\bibitem[\protect\astroncite{{Tafalla} et~al.}{2002}]{tafalla02}
{Tafalla}, M., {Myers}, P.~C., {Caselli}, P., {Walmsley}, C.~M., and {Comito},
  C.: 2002,
\newblock {\em \apj} {\bf 569}, 815

\bibitem[\protect\astroncite{{Tafalla} et~al.}{2006}]{tafalla06}
{Tafalla}, M., {Santiago-Garc{\'{\i}}a}, J., {Myers}, P.~C., {Caselli}, P.,
  {Walmsley}, C.~M., and {Crapsi}, A.: 2006,
\newblock {\em \aap} {\bf 455}, 577

\bibitem[\protect\astroncite{Testi et~al.}{1997}]{testietal97}
Testi, L., Palla, F., Prusti, T., Natta, A., and Maltagliati, S.: 1997,
\newblock {\em \aap} {\bf 320}, 159

\bibitem[\protect\astroncite{{Testi} and {Sargent}}{1998}]{testi98}
{Testi}, L. and {Sargent}, A.~I.: 1998,
\newblock {\em \apjl} {\bf 508}, L91

\bibitem[\protect\astroncite{{T{\'o}th} et~al.}{2004}]{toth04}
{T{\'o}th}, L.~V., {Haas}, M., {Lemke}, D., {Mattila}, K., and {Onishi}, T.:
  2004,
\newblock {\em \aap} {\bf 420}, 533

\bibitem[\protect\astroncite{{Townsley} et~al.}{2006}]{townsley06}
{Townsley}, L.~K., {Broos}, P.~S., {Feigelson}, E.~D., {Garmire}, G.~P., and
  {Getman}, K.~V.: 2006,
\newblock {\em \aj} {\bf 131}, 2164

\bibitem[\protect\astroncite{{Troland} and {Crutcher}}{2008}]{troland08}
{Troland}, T.~H. and {Crutcher}, R.~M.: 2008,
\newblock {\em ArXiv e-prints} 802

\bibitem[\protect\astroncite{{van der Werf}}{2000}]{vanderwerf00}
{van der Werf}, P.: 2000,
\newblock in F. {Combes} and G. {Pineau Des Forets} (eds.), {\em Molecular
  Hydrogen in Space}, p. 307

\bibitem[\protect\astroncite{{V{\'a}zquez-Semadeni} et~al.}{2003}]{Vazquez03}
{V{\'a}zquez-Semadeni}, E., {Ballesteros-Paredes}, J., and {Klessen}, R.~S.:
  2003,
\newblock {\em \apjl} {\bf 585}, L131

\bibitem[\protect\astroncite{{V{\'a}zquez-Semadeni}
  et~al.}{2009}]{2009ApJ...707.1023V}
{V{\'a}zquez-Semadeni}, E., {G{\'o}mez}, G.~C., {Jappsen}, A.,
  {Ballesteros-Paredes}, J., and {Klessen}, R.~S.: 2009,
\newblock {\em \apj} {\bf 707}, 1023

\bibitem[\protect\astroncite{{V{\'a}zquez-Semadeni}
  et~al.}{2007}]{2007ApJ...657..870V}
{V{\'a}zquez-Semadeni}, E., {G{\'o}mez}, G.~C., {Jappsen}, A.~K.,
  {Ballesteros-Paredes}, J., {Gonz{\'a}lez}, R.~F., and {Klessen}, R.~S.: 2007,
\newblock {\em \apj} {\bf 657}, 870

\bibitem[\protect\astroncite{V{\'a}zquez-Semadeni et~al.}{2009}]{vazsemetal09}
V{\'a}zquez-Semadeni, E., G{\'o}mez, G.~C., Jappsen, A.-K.,
  Ballesteros-Paredes, J., and Klessen, R.~S.: 2009,
\newblock {\em \apj} {\bf 707}, 1023

\bibitem[\protect\astroncite{{Veltchev} et~al.}{2011}]{veltchev11}
{Veltchev}, T.~V., {Klessen}, R.~S., and {Clark}, P.~C.: 2011,
\newblock {\em \mnras} {\bf 411}, 301

\bibitem[\protect\astroncite{{von Weizs\"{a}cker}}{1943}]{weizsaecker43}
{von Weizs\"{a}cker}, C.~F.: 1943,
\newblock {\em \za} {\bf 22}, 319

\bibitem[\protect\astroncite{{von Weizs{\"a}cker}}{1951}]{Weizsaecker51}
{von Weizs{\"a}cker}, C.~F.: 1951,
\newblock {\em \apj} {\bf 114}, 165

\bibitem[\protect\astroncite{{Walborn} and {Blades}}{1997}]{walborn97}
{Walborn}, N.~R. and {Blades}, J.~C.: 1997,
\newblock {\em \apjs} {\bf 112}, 457

\bibitem[\protect\astroncite{{Walder} and {Folini}}{1996}]{1996A&A...315..265W}
{Walder}, R. and {Folini}, D.: 1996,
\newblock {\em \aap} {\bf 315}, 265

\bibitem[\protect\astroncite{{Walder} and {Folini}}{1998}]{1998A&A...330L..21W}
{Walder}, R. and {Folini}, D.: 1998,
\newblock {\em \aap} {\bf 330}, L21

\bibitem[\protect\astroncite{{Walmsley} and {Ungerechts}}{1983}]{walmsley83}
{Walmsley}, C.~M. and {Ungerechts}, H.: 1983,
\newblock {\em \aap} {\bf 122}, 164

\bibitem[\protect\astroncite{{Wang} et~al.}{2010}]{wang10}
{Wang}, P., {Li}, Z., {Abel}, T., and {Nakamura}, F.: 2010,
\newblock {\em \apj} {\bf 709}, 27

\bibitem[\protect\astroncite{{Ward-Thompson}
  et~al.}{2007}]{WardThompsonetal2007}
{Ward-Thompson}, D., {Andr{\'e}}, P., {Crutcher}, R., {Johnstone}, D.,
  {Onishi}, T., and {Wilson}, C.: 2007,
\newblock in B. {Reipurth}, D. {Jewitt}, and K. {Keil} (eds.), {\em Protostars
  and Planets V}, pp 33--46

\bibitem[\protect\astroncite{{Ward-Thompson} et~al.}{2002}]{Ward02}
{Ward-Thompson}, D., {Andr{\'e}}, P., and {Kirk}, J.~M.: 2002,
\newblock {\em \mnras} {\bf 329}, 257

\bibitem[\protect\astroncite{{Ward-Thompson} et~al.}{1999}]{ward-thompson99}
{Ward-Thompson}, D., {Motte}, F., and {Andre}, P.: 1999,
\newblock {\em \mnras} {\bf 305}, 143

\bibitem[\protect\astroncite{{Weidner} and {Kroupa}}{2004}]{weidner04}
{Weidner}, C. and {Kroupa}, P.: 2004,
\newblock {\em \mnras} {\bf 348}, 187

\bibitem[\protect\astroncite{{Weidner} and {Kroupa}}{2006}]{weidner06}
{Weidner}, C. and {Kroupa}, P.: 2006,
\newblock {\em \mnras} {\bf 365}, 1333

\bibitem[\protect\astroncite{Weidner et~al.}{2010}]{weidetal10}
Weidner, C., Kroupa, P., and Bonnell, I.~A.~D.: 2010,
\newblock {\em \mnras} {\bf 401}, 275

\bibitem[\protect\astroncite{{Wilden} et~al.}{2002}]{wilden02}
{Wilden}, B.~S., {Jones}, B.~F., {Lin}, D.~N.~C., and {Soderblom}, D.~R.: 2002,
\newblock {\em \aj} {\bf 124}, 2799

\bibitem[\protect\astroncite{{Williams} et~al.}{2000}]{williams00}
{Williams}, J.~P., {Blitz}, L., and {McKee}, C.~F.: 2000,
\newblock {\em Protostars and Planets IV} p.~97

\bibitem[\protect\astroncite{{Williams} et~al.}{1994}]{Williamsetal1994}
{Williams}, J.~P., {de Geus}, E.~J., and {Blitz}, L.: 1994,
\newblock {\em \apj} {\bf 428}, 693

\bibitem[\protect\astroncite{Wolfire and Cassinelli}{1987}]{wolfcas87}
Wolfire, M.~G. and Cassinelli, J.~P.: 1987,
\newblock {\em \apj} {\bf 319}, 850

\bibitem[\protect\astroncite{{Wolfire} et~al.}{1995}]{wolfire95}
{Wolfire}, M.~G., {Hollenbach}, D., {McKee}, C.~F., {Tielens}, A.~G.~G.~M., and
  {Bakes}, E.~L.~O.: 1995,
\newblock {\em \apj} {\bf 443}, 152

\bibitem[\protect\astroncite{Wood and Churchwell}{1989}]{woodchurch89}
Wood, D.~O.~S. and Churchwell, E.: 1989,
\newblock {\em \apjs} {\bf 69}, 831

\bibitem[\protect\astroncite{{Yorke} and {Sonnhalter}}{2002}]{yorke02}
{Yorke}, H.~W. and {Sonnhalter}, C.: 2002,
\newblock {\em \apj} {\bf 569}, 846

\bibitem[\protect\astroncite{{Zinnecker} and {Yorke}}{2007}]{zinnecker07}
{Zinnecker}, H. and {Yorke}, H.~W.: 2007,
\newblock {\em \araa} {\bf 45}, 481

\end{thebibliography}
}

\end{document}